\documentclass[%
 reprint,
 amsmath,amssymb,
 aps,
pra,
]{revtex4-2}

\usepackage{graphicx}
\usepackage{dcolumn}
\usepackage{bm}
\usepackage{multirow}
\usepackage{graphicx}
\usepackage[table,xcdraw]{xcolor}
\usepackage{siunitx}    
\usepackage{ragged2e}
\usepackage{caption} 
\usepackage{booktabs}
\usepackage{hyperref}
\usepackage{booktabs}
\usepackage{mwe}
\usepackage{booktabs, makecell, multirow, tabularx}

\begin{document}

\title{Multi-channel OOK Communication by Using Frequency-multiplexed Switchable Metasurface}

\author{Parsa Farzin}
\email{parsa_farzin@elec.iust.ac.ir}
\affiliation{%
School of Electrical Engineering, Iran University of Science and Technology, Tehran, 1684613114, Iran.
}%
 
\author{Kasra Rouhi}%
\email{kasrarouhi@elec.iust.ac.ir}
\affiliation{%
School of Electrical Engineering, Iran University of Science and Technology, Tehran, 1684613114, Iran.
}%

\author{Seyed Ehsan Hosseininejad}%
\email{sehosseininejad@nit.ac.ir}
\affiliation{%
 Department of Electrical and Computer Engineering, Babol Noshirvani University of Technology, Babol, Iran.
}%

\begin{abstract}
Programmable metasurfaces have recently attracted considerable interest for their versatile applications in areas such as beam steering, holography, and wireless communications, utilizing either phase or amplitude modulation. Despite this, programmable amplitude coding modulation has seen limited exploration, primarily due to the difficulties involved in achieving real-time dynamic amplitude control. Here, we propose a reprogrammable amplitude-coding metasurface utilizing the on-off Keying (OOK) method combined with frequency modulation. To the best of our knowledge, this is the first time we will address both the design of the metasurface and the theoretical investigation of OOK simultaneously, considering all parameters present in the design, channel, and on-off ratio. The proposed metasurface comprises two layers of graphene with separate biasing voltages. By controlling the chemical potential of each layer, we can modulate the amplitude in two states at two frequencies through a field-programmable gate array (FPGA). In addition, we employ an information encryption method using the substitution cipher method and transmit it at two amplitude levels at distinct frequencies of $f_1 = 0.98\:\text{THz}$ and $f_2 = 1.46\:\text{THz}$ to safeguard transmission information against eavesdropping. Simulation and numerical results convincingly demonstrate that the proposed reprogrammable metasurface facilitates secure communication in multi-channel data encryption, terahertz (THz) data storage, information processing, and THz communication.

\end{abstract}

\maketitle


\section{Introduction}
Terahertz (THz) electromagnetic waves encompass the spectrum between infrared and microwave and range from 0.1 to 10 THz. This frequency range has versatile applications in indoor communication, inter-satellite communication, imaging, and biological detection and has attracted significant interest \cite{kleine2011review,lewis2014review}. Efforts to take advantage of THz waves have led to significant advancements in developing modern, multifunctional THz devices. In addition, the rapid advances in metamaterial technology are paving the way for highly capable THz platforms, marking a noteworthy progression in recent years. 

Metamaterials are artificial structures made from precisely arranged sub-wavelength periodic or non-periodic structures \cite{monticone2017metamaterial}. By systematically designing structures on the fundamental scale of materials, we can overcome the limitations imposed by certain natural laws and achieve remarkable functionalities that surpass the inherent properties of natural materials \cite{26}. They have resulted in significant breakthroughs in multiple fields, including radar, imaging, and wireless communications \cite{27,28}. However, metamaterials face considerable practical challenges due to the strong dispersion of resonant responses, lossy characteristics, and the complexities involved in manufacturing bulky three-dimensional structures \cite{29}.

As an alternative, two-dimensional (2D) counterparts of metamaterials, known as metasurfaces, have been extensively studied for their promising advantages, including low cost, compactness, and ease of fabrication \cite{hu2021review}. As long as the elements remain subwavelength in size, the principles governing metasurfaces can be applied across a wide spectrum of electromagnetic waves, from microwaves to visible light \cite{33}. Metasurfaces can control numerous electromagnetic phenomena, including converting plane waves to surface waves \cite{34}, lens \cite{35}, directional scattering \cite{36}, cloaking \cite{37}, beam steering \cite{jahangiri2024beam,farzin2024reprogrammable}, and holography \cite{38}. 
In a pioneering study, T. Cui et al. introduced the concept of digital metasurfaces, effectively linking the physical domain with the digital world. This advancement offers a new perspective, allowing metasurfaces to be reconsidered through information science \cite{cui2014coding}.
In the digital coding model, fundamental parameters of electromagnetic waves, such as amplitude \cite{40}, phase \cite{41}, polarization \cite{farzin2024multi}, and frequency \cite{ozgoli2024multichannel} can be represented as digital states \cite{44}. Consequently, unlike conventional metasurfaces, coding metasurfaces with well-defined elements can be programmed and encoded for real-time communication.

Recent extensive studies on amplitude modulation have greatly enriched the field of metasurface modulations \cite{45}. In \cite{62}, Jwair et al. designed an amplitude modulation metasurface capable of controlling the 1-bit amplitude in the microwave band using light-dependent resistors. 
In this case, the amplitude of the reflected wave is modulated between two states, "on" (bit "1") and "off" (bit "0"). In another study, Luo et al. advanced their research by modulating the amplitude of the reflected wave in six modes within the microwave band, instead of only two \cite{63}. As a result of this enhancement, the system may be able to transmit information at a higher capacity. Wu et al. and Hong et al. designed a metasurface capable of modulating amplitude across three distinct frequency bands within the microwave frequency \cite{64, 65}. Consequently, the number of available frequency channels was increased as a result of this innovation. The modulation of the amplitude across different frequencies not only enhances the number of channels and increases the transmission rate of information, but also increases the security of transmitted information when potential eavesdroppers are present. However, there is a very limited amount of research related to amplitude-frequency modulation metasurfaces in the THz frequency range compared to microwave frequency range.

On-off keying (OOK) is currently one of the most widely employed modulation techniques in wireless communication systems \cite{46}. In OOK modulation, the carrier wave amplitude switches between two distinct states, on and off, corresponding to the digital values "0" and "1" \cite{47}.
Additionally, in order to achieve a higher data rate in OOK modulation, high-speed amplitude modulation is essential \cite{48}. 
Microwave frequencies provide data transmission rates in the range of tens of gigabits per second, whereas the THz band can achieve capacities of several terabits per second \cite{66}.
Consequently, the THz band emerges as a crucial and practical solution to 6G wireless communication. To achieve this, active materials like graphene \cite{rouhi2019multi, tahmasebi2022parallel, 49}, vanadium dioxide \cite{john2020multipolar, kargar2020reprogrammable}, liquid crystal \cite{52}, and indium-tin-oxide (ITO) \cite{soleimani2022near} can be used since as the frequency increases toward the THz region, the pin-diode is not commercially available. Among the mentioned tunable materials, graphene stands out with its exceptionally high response speed, ranging from microseconds to femtoseconds and its unique electrical properties.

In this paper, we propose a reprogrammable amplitude-frequency modulation technique that utilizes OOK modulation to achieve dynamic reconfigurability and enhanced information security during transmission. To the best of our knowledge, this is the first time we have integrated metasurface design with the theoretical OOK method while reviewing all crucial parameters in design and channel modeling. 
For metasurface design, we employ two key sections responsible for amplitude and frequency modulation. In each section, we use a graphene square ring and graphene square patch to control both frequency and amplitude modulation, respectively. Each graphene layer is biased independently through a field-programmable gate array (FPGA), allowing precise control of the Fermi energy level. This configuration enables each graphene layer to control 1-bit amplitude and frequency modulation simultaneously and independently. As a result, we realize a 2-bit reconfigurable metasurface that adaptively controls amplitude and frequency for various applications. To protect against data loss due to eavesdropping, we implement an encryption scheme using a substitution cipher, encrypting information at two amplitude levels across two distinct frequency channels, $f_1 = 0.98 \: \text{THz}$ and $f_2 = 1.46 \: \text{THz}$. The encrypted data are transmitted securely via these channels, making it impossible for eavesdroppers to decrypt the information, even if intercepted. Our results demonstrate that this innovative rewritable information metasurface not only enhances communication security but also provides high flexibility for multi-channel information encryption, terahertz communication, data storage, and advanced information processing. This makes it a robust platform for the next generation of high-speed communication and secure wireless communication systems.

\begin{figure}
\centering
\includegraphics[width=1\columnwidth]{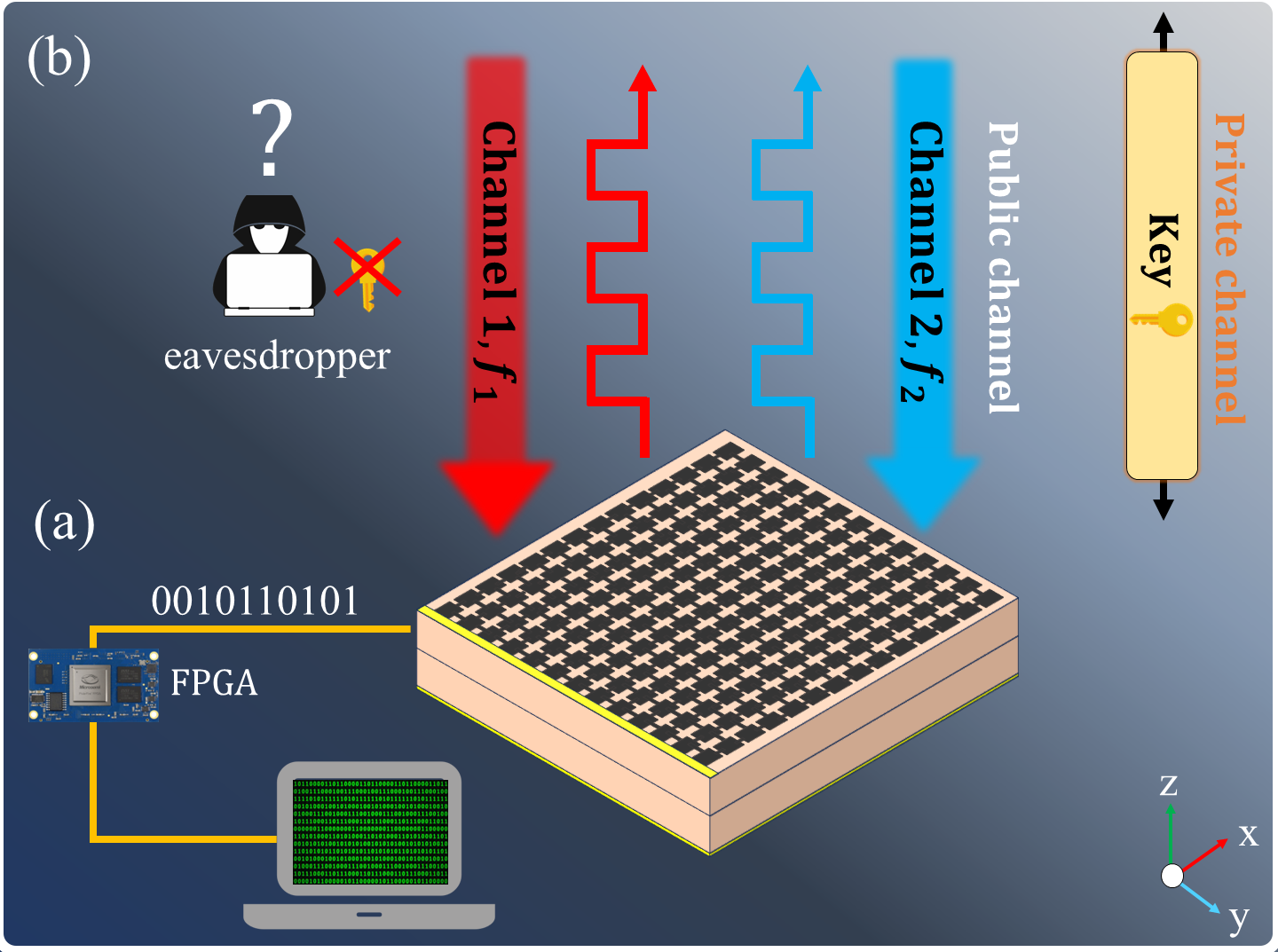}
\caption{The schematic of a reprogrammable amplitude-coding metasurface with frequency multiplexing for real-time THz wave communication encryption. (a) The information is encrypted using a substitution cipher algorithm, and the binary bits are processed through an FPGA for sequential configuration of the proposed metasurface. (b) The metasurface employs voltage-controlled amplitude modulation to transmit information over distinct frequencies on a public channel. The encrypted key is sent through a private channel accessible only to the sender and receiver, ensuring that an eavesdropper cannot decrypt the encrypted information even if it is intercepted.}
\label{fig: Main}
\end{figure}

\section{Communication Encryption Scheme Based on reprogrammable amplitude and frequency modulation Metasurface}
Figure \ref{fig: Main} provides a conceptual illustration of an encrypted wireless communication system that employs a reprogrammable metasurface to modulate both amplitude and frequency. In our proposed system, as illustrated in Fig. \ref{fig: Main}(a), the process begins on the sender’s side, where the target text is encrypted first using a substitution cipher. By replacing each character in plaintext with another character, this encryption method creates ciphertext that is resistant to interception and analysis by unauthorized parties. The binary information corresponding to the encrypted text is then provided to a field-programmable gate array (FPGA), a versatile hardware component capable of executing complex digital operations at high speed.

The FPGA serves as the controller that governs the operation of the metasurface by modulating both the amplitude and frequency of the reflected or transmitted signals. This modulation is performed by dynamically adjusting the bias voltages applied to the metasurface's graphene layers, which control the Fermi energy levels. Thus, the metasurface can switch between different states adaptively, encoding binary data as variations in amplitude and frequency. These variations allow for the transmission of information at two frequencies and amplitude levels, significantly increasing the data rate and making the system highly flexible for multi-channel communication.

It is important to note that the data is transmitted over a public communication channel, while the encryption key is sent through a separate private channel, accessible only to the sender and the authorized receiver, as shown in Fig. \ref{fig: Main}(b). Using this dual-channel approach, even if an adversary intercepts the public channel's data, they will not be able to decipher the information without access to the private key. The use of a substitution cipher further enhances the security of the system by complicating cryptanalysis attempts, as the transmitted data pattern is obscured by encryption.

The use of graphene as the primary material for the metasurface introduces several advantages, including high tunability, fast switching times, and the ability to operate across a wide range of frequencies, particularly in the THz spectrum. One of the key innovations in our design is the significant reduction in complexity compared to previous metasurface-based communication systems \cite{55,56,57,58,59,60,61}. Conventional controllable metasurfaces often require intricate biasing networks for each individual unit cell to achieve the desired electromagnetic response. These designs necessitate complex wiring and control systems, which increase both the physical footprint and the cost of the metasurface. In contrast, our proposed metasurface requires only a single set of external DC voltage sources to control the whole metasurface, which significantly simplifies the biasing architecture. As a result, our system is capable of modulating amplitude and frequency in real time, allowing for much faster transmission rates and more robust signal control.

\section{Materials and Methods}
\subsection{Complex Surface Conductivity of Graphene}
Graphene has attracted extensive interest for its remarkable characteristics as an almost transparent material with a zero-band gap, exhibiting semi-metallic properties. 
Due to graphene's sensitivity to external gate bias and its distinct frequency dispersion characteristics in the THz range for surface plasmon wave propagation, it serves as a highly effective material for dynamically controlling THz waves \cite{low2014graphene}. A graphene monolayer can be modeled as an ultra-thin carbon sheet with a complex surface impedance of $Z=1/\sigma_\mathrm{g}(\omega)$, where $\sigma_\mathrm{g}(\omega)$ represents the frequency-dependent complex conductivity of the material.
Applying Kubo's formula, the complex surface conductivity of graphene, distinguished between interband and intraband transitions, is given by \cite{hanson2008dyadic}

\begin{equation}
\sigma_{\mathrm{g}}(\omega) = \sigma^{\text{intra}}_{\text{g}}(\omega)+\sigma^{\text{inter}}_{\text{g}}(\omega),
\label{eq: GraphCond}
\end{equation}

\begin{equation}
\sigma_{\mathrm{g}}^{\mathrm{intra}}\left(\omega\right)=-\frac{ie^{2}k_{B}T}{\pi\hbar^{2}\left(\omega-i/\tau\right)}\left(\frac{\mu_{\mathrm{c}}}{k_{B}T}+2\ln\left(e^{\frac{-\mu_{\mathrm{c}}}{k_{B}T}}+1\right)\right),
\label{eq: Intra}
\end{equation}

\begin{equation}
\sigma_{\mathrm{g}}^{\mathrm{inter}}\left(\omega\right)=-\frac{ie^{2}}{4\pi\hbar}\ln\left[\frac{2\left|\mu_{c}\right|-\left(\omega-i\text{⁄}\tau\right)\hbar}{2\left|\mu_{c}\right|+\left(\omega-i\text{⁄}\tau\right)\hbar}\right],
\label{eq: Inter}
\end{equation}
where $k_\text{B}=1.38\times10^{-23}\:\mathrm{J/K}$ is the Boltzmann’s constant, $\hbar=h/2\pi$ represents the reduced Plack’s constant ($h = 6.62\times10^{-34}\:{\mathrm{J.s}}$ ), $e=1.6\times10^{-19}\:\mathrm{C}$ is electron charge, $\mu_\mathrm{c}$ is the chemical potential, $\tau$ is the electron-phonon relaxation time, and $T$ is the room temperature. In this paper all units are in the SI system, and the time variation is $\exp(i \omega t)$.
Here, we consider $\tau=0.6\:\mathrm{ps}$ and $T=300\:{\mathrm{K}}$. At room temperature and low THz frequencies, interband transitions can be neglected due to the Pauli exclusion principle, given that the photon energy is notably lower than the Fermi energy \cite{vasic2013tunable}. 
Controlling electromagnetic wave properties manipulated by graphene-based structures is achieved by modifying surface conductivity through external DC bias. This phenomenon can be quantified by changing the chemical potential $\mu_\mathrm{c}$ of graphene monolayer. The chemical potential can be dynamically tuned through real-time electrostatic biasing. Consequently, we can control electromagnetic waves by regulating external biasing using processors. The chemical potential of an isolated graphene sheet is dictated by the carrier density as below \cite{hanson2008dyadic},

\begin{figure}
\centering
\includegraphics[width=1\columnwidth]{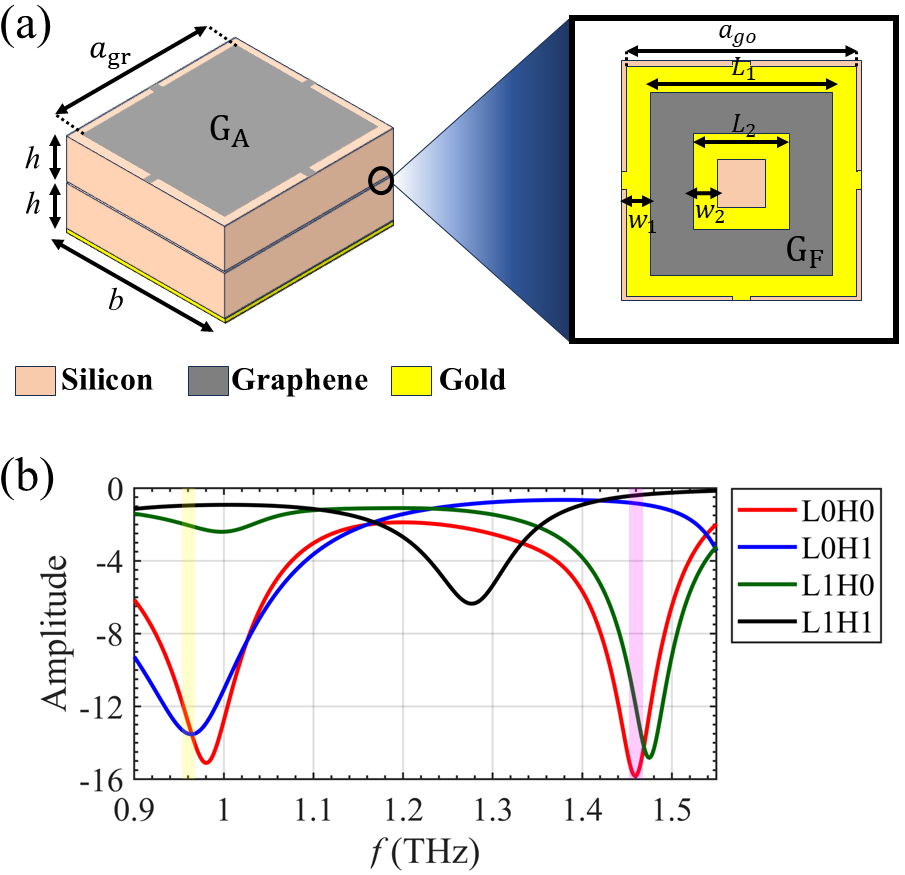}
\caption{(a) The proposed meta-atom design features two distinct graphene layers, each serving a specific modulation function. The top layer is dedicated to amplitude modulation, while the bottom layer handles frequency modulation. These layers are stacked on a silicon substrate, enabling real-time tunability of both amplitude and frequency by adjusting the chemical potentials of the graphene components. (b) The simulation results illustrate the 2-bit programmable modulation capability of the meta-atom, demonstrating amplitude and frequency control at two frequencies.}
\label{fig: UC}
\end{figure}

\begin{equation}
n_\mathrm{s} = \frac{2}{\pi\hbar^2\nu_\mathrm{F}^2} \int_{0}^{\infty} \epsilon(f_d (\epsilon)-f_d (\epsilon + 2\mu_c))  \,d\epsilon.
\label{eq: n_s}
\end{equation}
Here, $n_\mathrm{s}$ represents the 2D surface-charge density, $\epsilon$ is energy, and $\nu_\mathrm{F}$ is the Fermi velocity. Additionally, the Fermi-Dirac distribution is defined as \cite{hanson2008dyadic}

\begin{equation}
f_d(\epsilon)=(e^{(\epsilon-\mu_c)/k_\text{B} T)}+1)^{-1}
\label{eq: fermidirac}
\end{equation}
A straightforward expression of Eq. (\ref{eq: n_s}) reads $n_s = (\varepsilon_0 \varepsilon_\text{Aux} V_\text{DC})/(h_\text{Aux} e)$, where $\varepsilon_0$ and $\varepsilon_\text{Aux}$ are the permittivities of free space and material used in auxiliary layer required for biasing, respectively. Also, $V_\mathrm{DC}$ represents the necessary bias voltage and $h_\text{Aux}$ is the thickness of auxiliary  layer \cite{novoselov2004electric}. In this paper we assume that the auxiliary layer required for biasing is made of SiO2 with a thickness of $h_\text{Aux} = 300\:\text{nm}$. Also, the required chemical potential of the graphene sheet is given by \cite{ju2011graphene}

\begin{equation}
\mu_\mathrm{c} = \hbar \nu_\mathrm{F} \sqrt{\pi n_\mathrm{s}}.
\label{eq: BiasChemical}
\end{equation}
So, graphene's chemical potential can be dynamically altered by manipulating the applied DC voltage via capacitive biasing. The relationship between graphene's chemical potential and the gate voltage can be described as follows \cite{rouhi2021designing},

\begin{equation}
V_\mathrm{DC} = \frac{\mu_\mathrm{c} h_\mathrm{Aux} e}{ \hbar^2 \nu_\mathrm{F}^2 \pi \varepsilon_0 \varepsilon_\mathrm{Aux}}.
\label{eq: BiasVoltage}
\end{equation}

\subsection{Design Principle}
The key feature of the proposed metasurface is its ability to independently control the amplitude at one or two different frequencies, which can be achieved by choosing an appropriate chemical potential. To attain this objective, our design relies on a meta-atom comprising two key components: (i) amplitude modulation and (ii) frequency modulation. We employ two separate graphene components that control both amplitude and frequency. 
As shown in Fig. \ref{fig: UC}(a), the layers in the unit cell are stacked from bottom to top as follows: a silicon substrate ($\varepsilon_r=11.9$, $\text{tan}\delta=2.5×10^{-4}$) with a thickness of $h$ is placed on top of a thin gold layer, serving as the metallic layer. A graphene square ring $(\mathrm{G_F})$, a frequency modulation component, encircled by gold resonant rings, is positioned on this substrate and is enclosed by another layer of silicon substrate with a similar thickness on top. Finally, a graphene square patch $(\mathrm{G_A})$, amplitude modulation component, with a side length of $a_\text{gr}$ is placed on this substrate. 
In the proposed meta-atom, the periodic dimension of the unit cell is denoted as $b$ in both directions and the other geometric parameters are selected as, $h=25\:\mathrm{\mu m}$, $a_\text{gr}=45\:\mathrm{\mu m}$, $b=50\:\mathrm{\mu m}$, $L_1=40\:\mathrm{\mu m}$, $L_2=20\:\mathrm{\mu m}$, and $w_1=4.95\:\mathrm{\mu m}$, and $w_2=5\:\mathrm{\mu m}$.

As depicted in Fig. \ref{fig: UC}(a), the graphene segments responsible for amplitude and frequency control in the metasurface are connected by graphene and gold strips, respectively. This enables us to easily control the chemical potential of graphene in each part of the metasurface using a single set of electrostatic voltage, eliminating the need for independently adjusting each meta-atom. It is important to highlight that, for the electrostatic voltage bias applied to the graphene layer responsible for amplitude control, a gold electrical contact is linked to the graphene surface, as illustrated in Fig. \ref{fig: Main}.

As illustrated in \cite{liu2019simple}, incident wave amplitude modulation is exclusively achievable through a graphene patch with an appropriate chemical potential. Moving into the frequency modulation layer, graphene embedded between two resonant golden rings with a separate chemical potential, facilitates modifications to the structure's equivalent circuit, which leads to changing the operating frequency in real-time. This, in turn, enables real-time control over the structure's working frequency. This concept is analogous to the approach presented in \cite{11}, where altering the dimensions of the square rings results in a change in the working frequency. Hence, through the suitable design of the meta-atom and the utilization of a tunable material like graphene, we can control the amplitude of the incident wave independent of polarization in real-time at one or two different frequencies.
Utilizing distinct chemical potentials for amplitude and frequency modulation, enabling the performance of various functions, is depicted as $\mu_\mathrm{c} = [\mu_\mathrm{c}^{\mathrm{G_A}},\mu_c^{\mathrm{G_F}}]$, where $\mu_\mathrm{c}^{\mathrm{G_A}}$ denotes the chemical potential applied to $\mathrm{G_A}$(amplitude modulation layer), and $\mu_\mathrm{c}^{\mathrm{G_F}}$ represents the chemical potential applied to $\mathrm{G_F}$ (frequency modulation layer). By adjusting the graphene patch bias in the amplitude modulation layer, we are able to achieve amplitudes that correspond to both "0" and "1" bits in real-time.
The digital states "0" and "1" encode the low and high levels of the reflection coefficient amplitude in the amplitude coding system. Moreover, within the frequency modulation layer, the change of the proposed metasurface function from frequency $f_1$ to $f_2$ in real-time is achievable by adjusting the graphene ring bias. This transformation represents these two frequencies as digital states of bit "0" and bit "1", respectively. Therefore, through simultaneous and real-time control of amplitude at different frequencies, we can upgrade the metasurface control from 1-bit to 2-bit. Figure \ref{fig: UC}(b) illustrates the amplitude coefficients independent of polarization at two frequencies, $f_1=0.98\:\mathrm{THz}$ and $f_2=1.46\:\mathrm{THz}$.
We represent the 2-bit obtained from the amplitude control in both frequency channels as $\mathrm{L} b / \mathrm{H} b$, where "$\mathrm{L}$" and "$\mathrm{H}$" denote low frequency and high frequency, respectively. The parameter "$b$" indicates a bit that can be either 0 or 1 corresponding to low and high amplitude. To achieve control of absorption at both $f_1$ and $f_2$ frequencies, only absorption at $f_1$ frequency, only absorption at $f_2$ frequency, and reflection at both $f_1$ and $f_2$ frequencies, the chemical potential values for both $\mathrm{G_A}$ and $\mathrm{G_F}$ layers are denoted as $\mu_\text{c}=[0.3\:\text{eV}, 0.4\:\text{eV}]$, $\mu_\text{c}=[1\:\text{eV}, 0.2\:\text{eV}]$, $\mu_\text{c}=[0.1\:\text{eV}, 0.6\:\text{eV}]$ and $\mu_\text{c}=[0\:\text{eV}, 0\:\text{eV}]$, respectively. These values are assigned to the states "$\mathrm{L}0 / \mathrm{H}0$", "$\mathrm{L}0 / \mathrm{H}1$", "$\mathrm{L}1 / \mathrm{H}0$", and "$\mathrm{L}1 / \mathrm{H}1$", respectively.

\begin{figure}
\centering
\includegraphics[width=1\columnwidth]{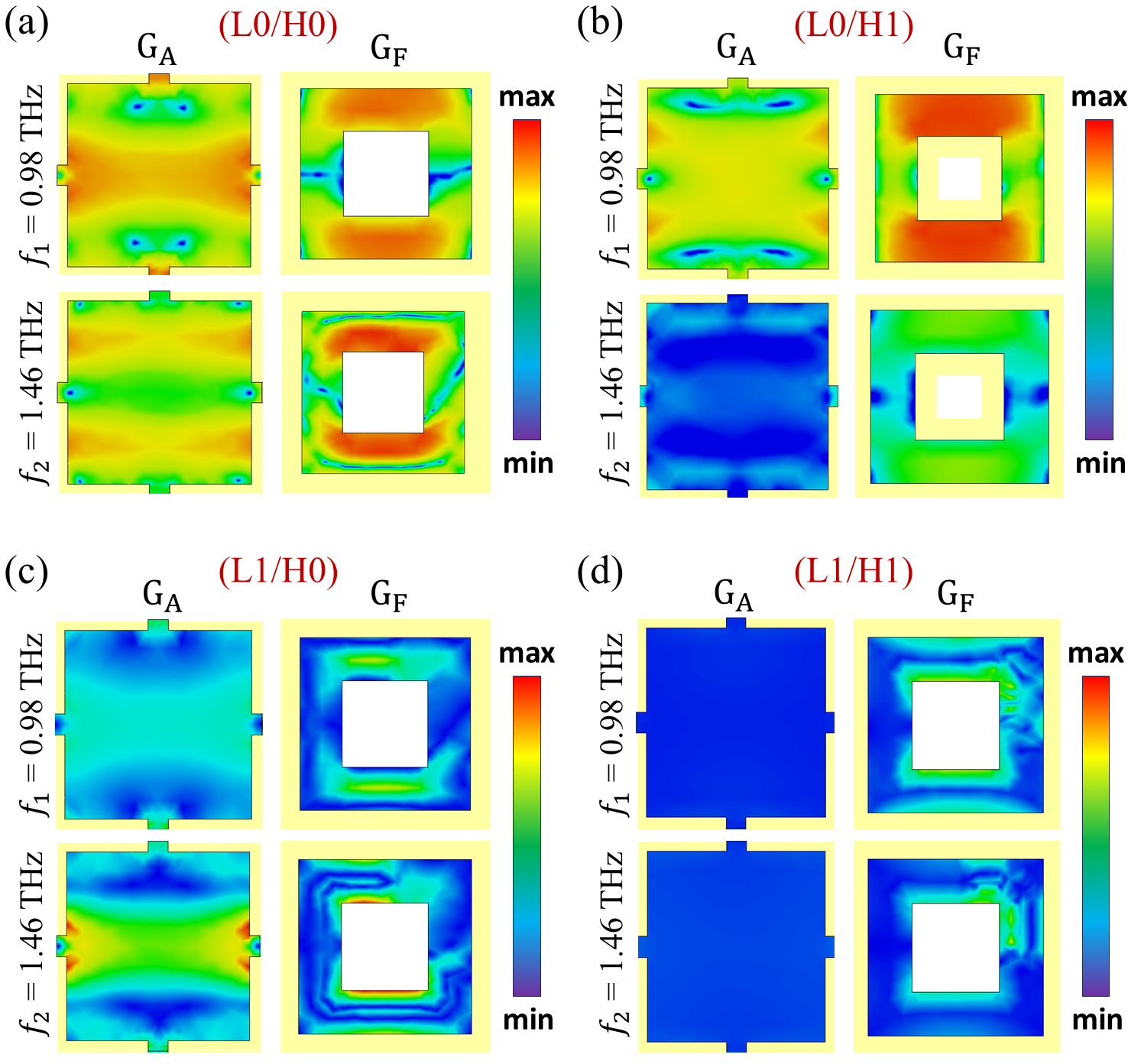}
\caption{The figure illustrates the simulated surface current density distributions for the meta-atom in various modulation states at two distinct terahertz frequencies, $f_1 = 0.98 \, \text{THz}$ and $f_2 = 1.46 \, \text{THz}$. 
(a) L0/H0 state: Both graphene layers exhibit strong surface current densities, indicating absorption at both frequencies $f_1$ and $f_2$.
(b) L0/H1 state: Strong surface current is observed at $f_1$, representing absorption, while $f_2$ shows weak surface current, indicating reflection. 
(c) L1/H0 state: The roles are reversed, with absorption occurring at $f_2$ and reflection at $f_1$, due to the modulated bias applied to the graphene layers. 
(d) L1/H1 state: Minimal surface current at both frequencies indicates reflection at $f_1$ and $f_2$. 
These current distributions demonstrate the dynamic tunability of the meta-atom, enabling real-time control over absorption and reflection at two distinct frequencies.
}
\label{fig:surface current}
\end{figure}

To clarify the mechanism of the proposed metasurface, the surface current density magnitude is also simulated. Figure \ref{fig:surface current} illustrates the current distribution for all four states to which the metasurface can switch. As depicted in Fig. \ref{fig:surface current}(a), in the $\mathrm{L}0 / \mathrm{H}0$ state (absorption at both $f_1$ and $f_2$ frequencies), both the $\mathrm{G_A}$ and $\mathrm{G_F}$ layers exhibit strong surface current density at both frequencies. This indicates that the electric field at both $f_1$ and $f_2$ frequencies is absorbed by the graphene in each layer. As shown in Fig. \ref{fig:surface current}(b), in the $\mathrm{L}0 / \mathrm{H}1$ state (absorption at the $f_1$ frequency and reflection at the $f_2$ frequency), the surface current density is strong at the $f_1$ frequency due to absorption, resulting in the electric field being absorbed in each layer. Conversely, at the $f_2$ frequency, where the structure is in reflection mode and no absorption occurs, the surface current density is weak, and the energy is almost fully reflected. In Fig. \ref{fig:surface current}(c), in the $\mathrm{L}1 / \mathrm{H}0$ state (reflection at frequency $f_1$ and absorption at frequency $f_2$), the absorption and reflection function is the opposite of the previous state, aligning with the obtained results. It should be noted that the relatively weak surface current density at $f=1.46\:\text{THz}$ in the $\mathrm{G_F}$ layer is due to its chemical potential being equal to 0.1 eV, causing the graphene to shift away from its absorption state. As shown in Fig. \ref{fig:surface current}(d), in the $\mathrm{L}1 / \mathrm{H}1$ state (reflection at both $f_1$ and $f_2$ frequencies), each graphene layer has 0 eV chemical potential and reflects the incident wave at both frequencies. Consequently, the current distribution density is very weak at both frequencies.

\begin{figure*}
\centering
\includegraphics[width=0.8\textwidth]{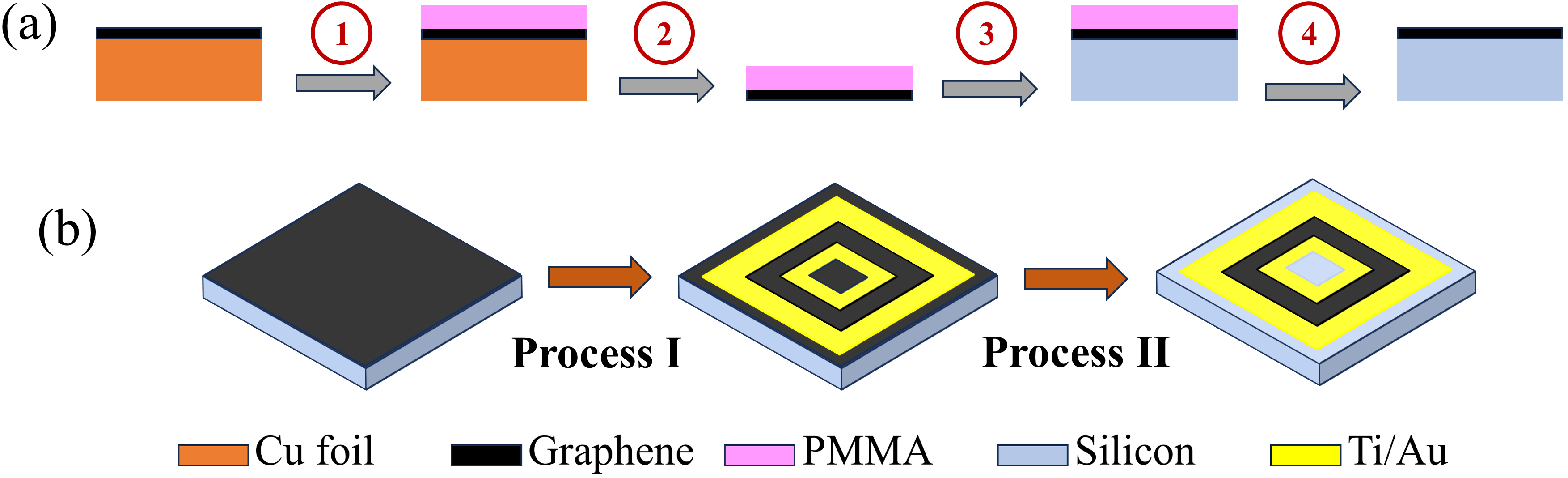}
\caption{The steps required for potential fabrication. (a) Graphene transfer flowcharts. (b) Fabrication process flow for graphene patterning and metal deposition.}
\label{fig:fabrication}
\end{figure*}

\subsection{Potentional Fabrication Steps}
To achieve cost-effective and large-scale production of single-layer graphene, chemical vapor deposition (CVD) takes center stage. This process essentially breaks down carbon atoms from hydrocarbons like methane, depositing them onto specific metal surfaces such as copper, ultimately resulting in the uniform and expansive formation of single-layer graphene \cite{12}. After graphene growth on copper foil within the CVD chamber, the single-layer graphene is transferred onto desired substrates using Polymethyl Methacrylate (PMMA) as a facilitating agent \cite{13}. As depicted in Fig. \ref{fig:fabrication}(a), the transfer process consists of four sequential steps: (i) initially, a PMMA layer is spin-coated atop the graphene layer; (ii) subsequently, the underlying copper support is etched away using a Cu etchant; (iii) next, the PMMA/graphene membrane is affixed onto the designated substrate, such as high-resistivity silicon; (iv) finally, the PMMA adhesive layer is dissolved in an organic solvent, leaving the graphene securely deposited onto the silicon. It's recommended to perform a distilled water rinse to remove any chemical residues that may remain during the transfer process. Additionally, enhancing the adhesion between graphene and the substrate can be achieved by baking for several hours, allowing complete evaporation of water molecules \cite{12}. Hence, the square patch of graphene ($\mathrm{G_A}$) is fabricated in this manner. However, to create the graphene ring ($\mathrm{G_F}$) encircled by two gold rings, we employ fabrication techniques commonly used for electrode deposition and graphene patterning. As illustrated in Fig. \ref{fig:fabrication}(b), the process of depositing coplanar electrodes with two rings (Process I) comprises three stages: (i) utilizing photo-lithography to define the pattern of two coplanar gold rings on the photoresist layer; (ii) deposition of a bi-metal layer (20 nm Ti/100 nm Au) onto the graphene sheet using e-beam evaporation; (3) Removal of unwanted metal during the lift-off process \cite{12}. Graphene patterning (Process II) is accomplished through three steps: (i) applying a photoresist pattern covering an area of $L_1 \times L_1$ $\mathrm{\mu m^2}$ on the graphene sheet between the two gold rings; (ii) Etching away the graphene sheet not covered by the photoresist using oxygen plasma, it's important to note that the plasma power must not be too high to avoid damaging the protective photoresist; (iii) stripping the photoresist using acetone followed by a distilled water rinse \cite{12}. 

\section{On-off-keying for Free-space Optical Communication System}
\subsection{Channel Model}
\subsubsection{Log-normal Distribution}
As electromagnetic waves travel through an atmospheric turbulence channel, their intensity experiences fluctuations due to scintillation. Several models have been proposed to characterize scintillation in such environments \cite{wang1974perturbed, andrews2005laser, letzepis2009outage}. Among these, the log-normal distribution is frequently used to describe scintillation under weak to moderate turbulence conditions \cite{andrews2005laser, nistazakis2009performance}.
In the log-normal distributed channel model, the probability density function (PDF) of the normalized wave intensity $I_\mathrm{S}$ is expressed by \cite{osche2002optical, popoola2008free, kwok2008link, wang2012performance}

\begin{equation}
f_{\mathrm{LN}}\left(I_{\mathrm{S}}\right)=\frac{1}{\sqrt{2\pi\sigma_{\mathrm{R}}^{2}}I_{\mathrm{S}}}\exp\left\{ -\frac{\left[\ln\left(\frac{I_{\mathrm{S}}}{I_{0}}\right)+\frac{\sigma_{\mathrm{R}}^{2}}{2}\right]^{2}}{2\sigma_{\mathrm{R}}^{2}}\right\},	\label{eq: f_LN}
\end{equation}
where $\sigma_\mathrm{R}^2$ is the Rytov variance, which is a general parameter representing the strength of atmospheric turbulence and $I_0$ denotes the received intensity without turbulence, which is assumed to be unity \cite{osche2002optical}. Then, the Rytov variance is expressed by \cite{osche2002optical, andrews2005laser, popoola2008free, wang2012performance}
\begin{equation}
	\sigma_\mathrm{R}^2=1.23 C_\mathrm{n}^2 k^{\frac{7}{6}} L^{\frac{11}{6}},
\end{equation}
where $L$ is the transmission distance, which is the total distance the wave travels through the turbulent medium.
Also, $k=2 \pi / \lambda$ is the wavenumber and $\lambda$ is the wavelength of the propagated wave.
$C_\mathrm{n}^2$ is refractive index structure constant, which characterizes the strength of turbulence in the atmosphere and varying from $10^{-17} \mathrm{~m}^{-2/3}$ to $10^{-13} \mathrm{~m}^{-2/3}$. 
It measures the strength of the turbulence-induced fluctuations in the refractive index of the medium (such as air) and larger values of this parameter indicate stronger turbulence, which leads to more significant fluctuations in the phase and amplitude of the wave.
Different Rytov variance values represent different atmospheric turbulence conditions. So, the weak, moderate and strong intensity fluctuations are associated with $\sigma_\mathrm{R}^2<1, \sigma_\mathrm{R}^2 \approx 1$ and $\sigma_\mathrm{R}^2>1$, respectively \cite{andrews2005laser}.

\subsubsection{Gamma-Gamma Distribution}
As discussed in \cite{osche2002optical, popoola2008free}, the log-normal distribution becomes invalid when $\sigma_\mathrm{R}^2 > 1.2$, which occurs in the strong scintillation regime where significant intensity fluctuations and deep fading are observed. In such cases, the Gamma-Gamma distribution is often employed to model strong turbulence conditions \cite{letzepis2009outage}, where $\sigma_\mathrm{R}^2 > 1$.  In this approach, the intensity of the received signal is modeled as the product of two independent Gamma-distributed random variables. Using the Gamma-Gamma distribution, which is a composite model, the PDF of $I_\mathrm{S}$ can be derived as \cite{al2001mathematical, andrews2005laser, li2017suboptimal},
\begin{equation}
	f_\mathrm{GG}\left(I_\mathrm{S}\right)=\frac{2(\alpha \beta)^{\frac{\alpha+\beta}{2}}}{\Gamma(\alpha) \Gamma(\beta)} I_\mathrm{S}^{\frac{\alpha+\beta}{2}-1} K_{\alpha-\beta}\left(2 \sqrt{\alpha \beta I_\mathrm{S}}\right),
	\label{eq: f_GC}
\end{equation}
where $\Gamma$ denotes the Gamma function, and $K_n(\cdot)$ represents the modified Bessel function of the second kind of order $n$. The parameters $\alpha$ and $\beta$ are shape factors that correspond to small-scale and large-scale turbulence effects, respectively.
They used in the PDF and they are defined by Rytov variance $\sigma_\mathrm{R}^2$, which are given by \cite{al2001mathematical, li2017suboptimal}
\begin{equation}
\alpha\cong\left[\exp\left[\frac{0.49\sigma_{\text{R}}^{2}}{\left(1+1.11\sigma_{\text{R}}^{\frac{12}{5}}\right)^{\frac{7}{6}}}\right]-1\right]^{-1},
\end{equation}

\begin{equation}
\beta\cong\left[\exp\left[\frac{0.51\sigma_{\text{R}}^{2}}{\left(1+0.69\sigma_{\text{R}}^{\frac{12}{5}}\right)^{\frac{7}{6}}}\right]-1\right]^{-1}.
\end{equation}
Here, $\alpha$ describes the effect of small-scale turbulence, which corresponds to rapid intensity fluctuations caused by small eddies or short-term changes in the refractive index and $\beta$ describes the effect of large-scale turbulence, associated with slower, longer-term variations in intensity caused by larger eddies in the atmosphere.
In summary, the values of $\alpha$ and $\beta$ depend on the Rytov variance and the specific turbulence conditions of the environment, typically ranging from about 1 to 10, depending on the turbulence strength.

\subsection{Signal Model}
We consider long-distance intensity modulation with direct detection (IM/DD) link using the OOK modulation system. This is a simple and widely-used method in optical communication systems because it avoids the complexity of more advanced techniques like coherent detection. A high degree of accuracy can also be achieved by modeling additive noise as additive white Gaussian noise (AWGN), which is statistically independent of the transmitted data and intensity scintillation \cite{zhu2002free}. The parameter $\tau_\text{s}$ denotes the bit interval of the sequence of OOK-modulated data and for the $m$-th bit interval, the discrete-time signal $r[m]$ is acquired by sampling the integral output of a detector at time $m \tau_\text{s}$. The received signal at the receiver can be expressed as \cite{zhu2002free}
\begin{equation}
	r[m]=s[m] I_s+i_n[m],
\end{equation}
where $s[m]$ denotes the transmitted datum, which can be either "0" or "1". Also, $I_\mathrm{s}$ is the channel fading intensity caused by atmospheric turbulence and $i_\mathrm{n}[m]$ is AWGN with variance $\sigma_\mathrm{n}^2$ and mean zero. As a result, the signal-to-noise ratio (SNR) can be expressed as follows:
\begin{equation}
\gamma=\frac{E\left[\left(s[m]I_{\mathrm{s}}\right)^{2}\right]}{E\left[\left(i_{\mathrm{n}}[m]\right)^{2}\right]}=\frac{E\left[\left(s[m]I_{\mathrm{s}}\right)^{2}\right]}{\sigma_\mathrm{n}^{2}},
\end{equation}
where $E[.]$ is the statistical expectation function. We presume that the channel state information (CSI) is not available at the receiver side. Hence, the PDF of the received signal $r[m]$ is \cite{proakis2008digital, borah2009pointing}
\begin{equation}
	P(r)=P(0) \cdot P(r|0)+P(1) \cdot P(r|1),
	\label{eq: P}
\end{equation}
where $P(0)$ and $P(1)$ represent the probabilities of transmitting "0" and "1" respectively, and for simplicity, we assume $P(0) = P(1) = 0.5$. Additionally, $P(r|0)$ and $P(r|1)$ denote the conditional error probabilities for the transmission of "0" and "1".
Since scintillation is a multiplicative effect, it predominantly impacts the transmission of "1".  It means that scintillation, which multiplies the signal intensity, mostly impacts the bits where the signal is "on" (data "1"), because this bit relies on a stronger field intensity, making it more susceptible to intensity fluctuations caused by scintillation. The conditional error probability is computed as \cite{li2007optical, borah2009pointing}
\begin{equation}
	P(r|0)=\int_{I_\mathrm{th}}^{\infty} \frac{1}{\sqrt{2 \pi \sigma_\mathrm{n}^2}} \exp \left(-\frac{r^2}{2 \sigma_\mathrm{n}^2}\right) d r=Q\left(\frac{I_\mathrm{th}}{\sigma_\mathrm{n}}\right),
	\label{eq: P0}
\end{equation}

\begin{equation}
\begin{aligned}
P(r|1) = \int_{-\infty}^{\infty} \int_{-\infty}^{I_\mathrm{th}} \frac{1}{\sqrt{2 \pi \sigma_\mathrm{n}^2}} \exp \left(-\frac{(r-I)^2}{2 \sigma_\mathrm{n}^2}\right) f(I) dr dI \\
=\int_{-\infty}^{\infty} Q\left(\frac{I-I_\mathrm{th}}{\sigma_\mathrm{n}}\right) f(I) d I.
\end{aligned}
\label{eq: P1}
\end{equation}
Here, the detector's responsivity is assumed to be unity, $Q$ is the $Q$-function, $r$ is the detected signal, $I$ represents the normalized intensity $I_\mathrm{S}$ and $f(I)$ is the PDF of $I_\mathrm{S}$ \cite{agrawal2012fiber}. 
By minimizing the bit error rate (BER), the decision threshold $I_\mathrm[th]$ is determined by the likelihood function $\Lambda$ at the unity value \cite{zhu2002free, borah2009pointing}.
In \cite{zhu2002free}, a maximum likelihood symbol-by-symbol detection (MLD) is proposed using the marginal distribution of the turbulence-induced fading. 
This method helps detect individual transmitted symbols by selecting the one most likely to match the received signal, accounting for the known effects of turbulence-induced fading. The conditional probability density function (CPDF) when transmitted data are "0" and "1" is given in Eqs. (\ref{eq: P0}) and (\ref{eq: P1}). The likelihood ratio for MLD is given by
\begin{equation}
\begin{aligned}
\Lambda & = \frac{P(r|1)}{P(r|0)} = 
\frac{\int_{-\infty}^{\infty} \frac{1}{\sqrt{2 \pi \sigma_\mathrm{n}^2}} \exp \left(-\frac{\left(I_\mathrm{th}-I\right)^2}{2 \sigma_\mathrm{n}^2}\right) f(I) \: dI} {\left(\frac{1}{\sqrt{2 \pi \sigma_\mathrm{n}^2}} \exp \left(-\frac{I_\mathrm {th}^2}{2 \sigma_\mathrm{n}^2}\right)\right)}.
\end{aligned}
\label{eq: Lambda}
\end{equation}
The likelihood ratio grows monotonically with $r$. 
Also, MLD can be achieved by comparing the received signal $r$ with the threshold value of $r_\text{th}$, which is the result of $\Lambda(r_\text{th})=1$ or $P(r_\text{th}|0)=P(r_\text{th}|1)$.
According to the above equations, the threshold $r_\text{th}$ is determined by information about the channel model. 
The optimum detection threshold versus turbulence strength is shown in Fig. 3 of \cite{zhu2002free}, in which this value is affected by AWGN variance $\sigma_\text{n}^2$ and the turbulence strength $\sigma_\text{R}^2$.
Substituting $I_\mathrm{th}$ into Eqs. (\ref{eq: P0}) and (\ref{eq: P1}), and then Eqs. (\ref{eq: P0}) and (\ref{eq: P1}) into Eq. (\ref{eq: P}), the BER performance of OOK signal will be
\begin{equation}
	P_{\mathrm{MLD}}(r)=\frac{1}{2}\left(Q\left(\frac{I_\mathrm{th}}{\sigma_n}\right)+\int_{-\infty}^{\infty} Q\left(\frac{I-I_\mathrm{th}}{\sigma_n}\right) f(I) d I\right) .
\end{equation}

We can obtain BERs before and after signal processing by replacing $f(I)$ in Eq. (\ref{eq: P1}) with $f(I_\mathrm{S})$ in Eq. (\ref{eq: f_LN}) for log-normal distribution or in Eq. (\ref{eq: f_GC}) for Gamma-Gamma distribution.
The method cannot be applied without knowing the channel model information. In the absence of channel information, suboptimal maximum likelihood detection (SMLD) is proposed. SMLD is a method proposed for use when channel information is unavailable. It allows symbol detection by determining a suboptimal threshold from the received signal $r$ without relying on the channel model, making it a practical alternative to traditional MLD in such situations. We know $P(r|0)$ is an even function; so

\begin{equation}
P(r|0)=P(-r|0).
\label{eq: P_neg}
\end{equation}
Since we assumed the probability of "0" and "1" is equal to 0.5; so, Eq. (\ref{eq: P}) is rewritten as
\begin{equation}
P(r) = 0.5 \cdot P(r|0) + 0.5 \cdot P(r|1).
\label{eq: P_eq}
\end{equation}
Given that the threshold is greater than zero in Eq. (\ref{eq: P_eq}), we set $r = r_\text{th}$ and substitute Eq. (\ref{eq: P_neg}) into this expression,
\begin{equation}
\begin{aligned}
P(r_\text{th}) & = 0.5 \cdot P(r_\text{th}|0) + 0.5 \cdot P(r_\text{th}|1) \\
& = 0.5 \cdot P(r_\text{th}|0) + 0.5 \cdot P(r_\text{th}|0) = P(r_\text{th}|0),
\end{aligned}
\label{eq: PP}
\end{equation}
Then, in Eq. (\ref{eq: P_eq}), we put $r=-r_\text{th}$,
\begin{equation}
P(-r_\text{th}) = 0.5 \cdot P(-r_\text{th}|0) + 0.5 \cdot P(-r_\text{th}|1).
\label{eq: P_neg2}
\end{equation}

There is a high probability that the received signal $r$ will be larger than 0 if the transmitted bit is "1", unless there is an extremely large amount of additive noise or deep fading. 
Additionally, the probability of $r>0$ is equal to the probability of $r<0$ when the transmitted bit is "0" due to the AWGN being considered and we presume that $P(r|0)$ is much bigger than $P(r|1)$ when $r<0$. Then, $P(-r_\text{th}|0) \gg P(-r_\text{th}|1)$, and by substituting Eq. (\ref{eq: P_neg}) into Eq. (\ref{eq: P_neg2}), we have

\begin{equation}
	P(-r_\text{th}) \approx 0.5 \cdot P(-r_\text{th}|0) = 0.5 \cdot P(r_\text{th}|0) .
	\label{eq: P_neg3}
\end{equation}
Comparing Eq. (\ref{eq: PP}) with Eq. (\ref{eq: P_neg3}), the suboptimal threshold can be obtained by $P\left(r_\mathrm{th}\right)=2 \cdot P\left(-r_\mathrm{th}\right)$, where $P(\cdot)$ shows the PDF of the received signal. As a result, the suboptimal threshold is determined without any channel information.

\begin{figure*}
	\centering
	\includegraphics[width=0.8\textwidth]{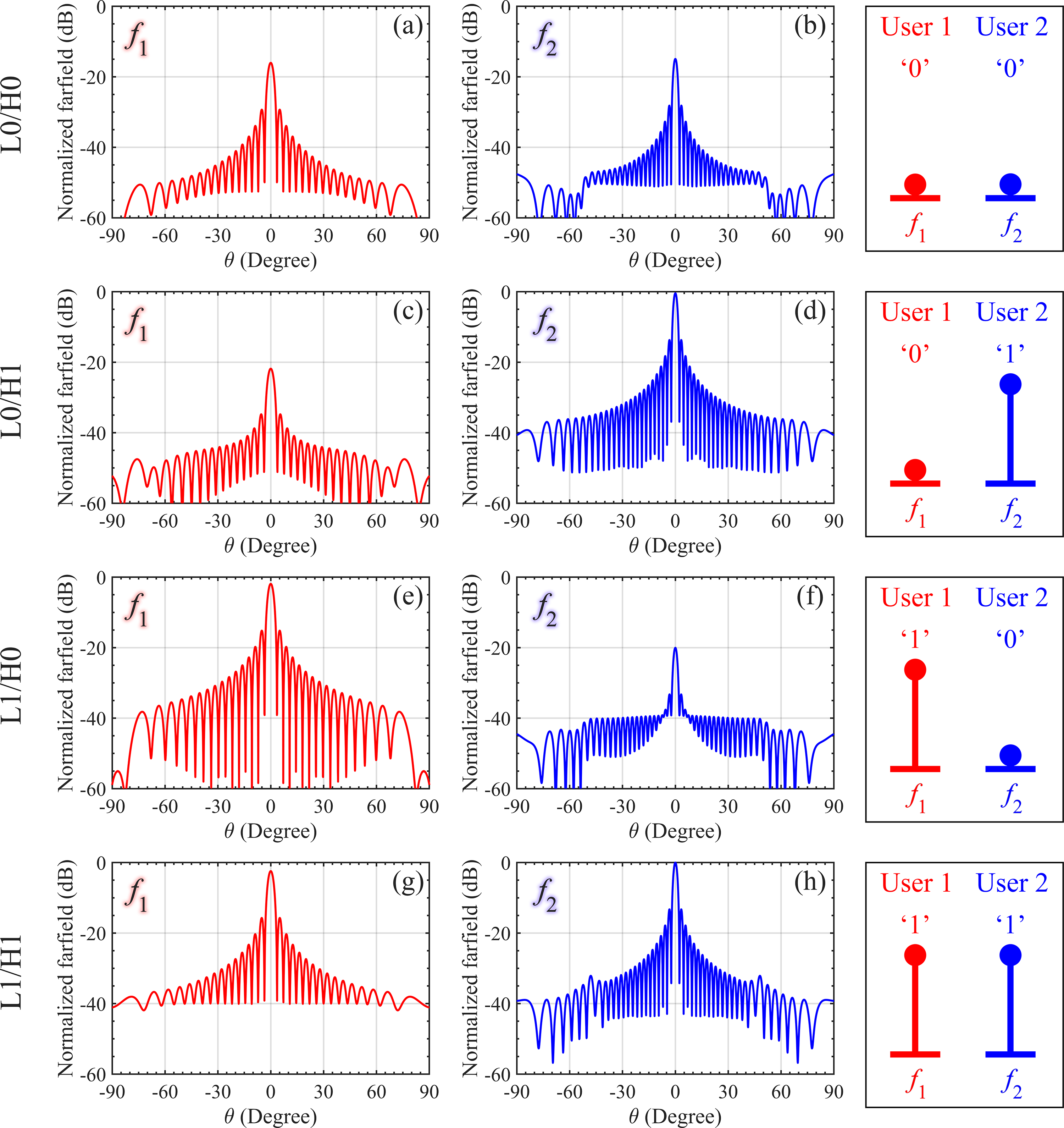}
	\caption{Scattering pattern controls of the metasurface and the spectral distributions received independently by users 1 and 2 are characterized by high power representing the binary symbol '1' and low power representing the binary symbol '0' in the OOK scheme. (a) and (b) the scattering patterns for $\mathrm{L}0 / \mathrm{H}0$ state at channel f1 and f2, respectively. (c) and (d) the scattering patterns for $\mathrm{L}0 / \mathrm{H}1$ state at channel f1 and f2, respectively. (e) and (f) the scattering patterns for $\mathrm{L}1 / \mathrm{H}0$ state at channel f1 and f2, respectively. (g) and (h) the scattering patterns for $\mathrm{L}1 / \mathrm{H}1$ state at channel f1 and f2, respectively.}
	\label{fig: Scattering}
\end{figure*}

\subsection{On-Off Ratio}
The on-off ratio of the OOK modulator is also another significant parameter which affects the system BER and is defined by

\begin{equation}
	r_\mathrm{OOK}=\frac{\Gamma_\mathrm{high}}{\Gamma_\mathrm{low}},
\end{equation}
where $\Gamma_\mathrm{high}$ and $\Gamma_\mathrm{low}$ are the reflection amplitude at the "on" (bit "1") and "off" (bit "0") states, respectively. 
It is ideal for $r_\mathrm{OOK}$ to be infinite since $\Gamma_\mathrm{low}$ equals 0. Any leakage within the modulator circuit, however, may result in a nonzero amplitude at the "off" state, and a higher transmitter output power is required in order to keep the same BER. It is estimated that in the worst case scenario, $r_\mathrm{OOK}=4.2$ will occur in our developed design.

\begin{table*}[t]
	\caption{Substitution cipher method using symbols to encrypt data and enhance the security of transmitted information.}
	\centering
	\label{tab: Cipher}
	\setlength{\arrayrulewidth}{0.4mm}
	\scalebox{1.1}{%
		\begin{tabular}{|l|c|c|c|c|c|c|c|c|c|c|c|c|c|c|c|}
			\hline
			\rowcolor[gray]{0.9} 
			\cellcolor[gray]{0.7}\rule{0pt}{10pt}\textbf{Plaintext alphabet/symbol} & \rule{0pt}{10pt}\textbf{a} & \rule{0pt}{10pt}\textbf{b} & \rule{0pt}{10pt}\textbf{c} & \rule{0pt}{10pt}\textbf{d} & \rule{0pt}{10pt}\textbf{e} & \rule{0pt}{10pt}\textbf{f} & \rule{0pt}{10pt}\textbf{g} & \rule{0pt}{10pt}\textbf{h} & \rule{0pt}{10pt}\textbf{i} & \rule{0pt}{10pt}\textbf{j} & \rule{0pt}{10pt}\textbf{k} & \rule{0pt}{10pt}\textbf{l} & \rule{0pt}{10pt}\textbf{m} & \rule{0pt}{10pt}\textbf{n} & \rule{0pt}{10pt}\textbf{o} \\ \hline
			\rowcolor[gray]{1.0} 
			\textbf{Ciphertext alphabet/symbol} & \rule{0pt}{10pt}t & \rule{0pt}{10pt}1 & \rule{0pt}{10pt}a & \rule{0pt}{10pt}s & \rule{0pt}{10pt}( & \rule{0pt}{10pt}j & \rule{0pt}{10pt}l & \rule{0pt}{10pt}p & \rule{0pt}{10pt}f & \rule{0pt}{10pt}b & \rule{0pt}{10pt}e & \rule{0pt}{10pt}! & \rule{0pt}{10pt}4 & \rule{0pt}{10pt}2 & \rule{0pt}{10pt}d \\ \hline
			\rowcolor[gray]{0.9} 
			\cellcolor[gray]{0.7}\rule{0pt}{10pt}\textbf{Plaintext alphabet/symbol} & \rule{0pt}{10pt}\textbf{p} & \rule{0pt}{10pt}\textbf{q} & \rule{0pt}{10pt}\textbf{r} & \rule{0pt}{10pt}\textbf{s} & \rule{0pt}{10pt}\textbf{t} & \rule{0pt}{10pt}\textbf{u} & \rule{0pt}{10pt}\textbf{v} & \rule{0pt}{10pt}\textbf{w} & \rule{0pt}{10pt}\textbf{x} & \rule{0pt}{10pt}\textbf{y} & \rule{0pt}{10pt}\textbf{z} & \rule{0pt}{10pt}\textbf{0} & \rule{0pt}{10pt}\textbf{1} & \rule{0pt}{10pt}\textbf{2} & \rule{0pt}{10pt}\textbf{3} \\ \hline
			\rowcolor[gray]{1.0} 
			\textbf{Ciphertext alphabet/symbol} & \rule{0pt}{10pt}3 & \rule{0pt}{10pt}7 & \rule{0pt}{10pt}m & \rule{0pt}{10pt}? & \rule{0pt}{10pt}z & \rule{0pt}{10pt}g & \rule{0pt}{10pt}5 & \rule{0pt}{10pt}o & \rule{0pt}{10pt}n & \rule{0pt}{10pt}= & \rule{0pt}{10pt}/ & \rule{0pt}{10pt}c & \rule{0pt}{10pt}- & \rule{0pt}{10pt}y & \rule{0pt}{10pt}k \\ \hline
			\rowcolor[gray]{0.9} 
			\cellcolor[gray]{0.7}\rule{0pt}{10pt}\textbf{Plaintext alphabet/symbol} & \rule{0pt}{10pt}\textbf{4} & \rule{0pt}{10pt}\textbf{5} & \rule{0pt}{10pt}\textbf{6} & \rule{0pt}{10pt}\textbf{7} & \rule{0pt}{10pt}\textbf{8} & \rule{0pt}{10pt}\textbf{9} & \rule{0pt}{10pt}\textbf{+} & \rule{0pt}{10pt}\textbf{=} & \rule{0pt}{10pt}\textbf{-} & \rule{0pt}{10pt}\textbf{(} & \rule{0pt}{10pt}\textbf{)} & \rule{0pt}{10pt}\textbf{?} & \rule{0pt}{10pt}\textbf{!} & \rule{0pt}{10pt}\textbf{@} & \rule{0pt}{10pt}\textbf{/} \\ \hline
			\rowcolor[gray]{1.0} 
			\textbf{Ciphertext alphabet/symbol} & \rule{0pt}{10pt}+ & \rule{0pt}{10pt}x & \rule{0pt}{10pt}h & \rule{0pt}{10pt}u & \rule{0pt}{10pt}0 & \rule{0pt}{10pt}@ & \rule{0pt}{10pt}8 & \rule{0pt}{10pt}v & \rule{0pt}{10pt}9 & \rule{0pt}{10pt}q & \rule{0pt}{10pt}w & \rule{0pt}{10pt}r & \rule{0pt}{10pt}i & \rule{0pt}{10pt}) & \rule{0pt}{10pt}6 \\ \hline
		\end{tabular}
	}
\end{table*}

\subsection{OOK Modulation Scheme}
We outline the process for simulating an OOK modulation scheme with a graphene-based modulator in this section. 
To accurately replicate the transmission of information through binary amplitude-shift keying (BASK), the frequency-dependent nature of the channel $S_{12}(f)$ and the linearity of the wave equation were leveraged. 
The data stream, consisting of $N$ bits, was represented as a $1 \times N$ vector $\mathbf{b}_{\mathrm{t}}$, with the $i$-th element of the vector indicated as $\mathbf{b}_{\mathrm{t}}[i]$.
Firstly, the transmitted signal $s_{\mathrm{TX}}(t)$ was generated by modulating the amplitude of a sinusoidal carrier at the modulation frequency $f_0$ using the data stream $\mathbf{b}_{\mathrm{t}}$, with a predefined bit duration of $\tau_\text{s}$:

\begin{equation}
\left.s_{\mathrm{TX}}\left(t\right)\right|_{\left(i-1\right)\tau_{\text{s}}\leq t<i\tau_{\text{s}}}=\left\{ \begin{array}{cl}
0 & \text{ if }\mathbf{b}_{\mathrm{t}}[i]=0\\
A\sin\left(2\pi f_{0}t\right) & \text{ if }\mathbf{b}_{\mathrm{t}}[i]=1
\end{array}\right.
\end{equation}
Here, the modulation speed is determined as $1 / \tau_\text{s}$ samples per second and the carrier amplitude $A$ specifies the transmitted power $P_{\mathrm{t}}=10 \log {10}\left((A / 2)^2 / 10^{-3}\right) \text{dBm}$, assuming both defined data symbols are equiprobable. It is also important to note that $s_{\mathrm{TX}}(t)$ should be strongly oversampled in order to avoid sampling errors.

Secondly, the received signal is determined as $s_{\mathrm{RX}}(t)$ and the Fourier transform yields $S_{\mathrm{RX}}(f)$. Then, $S_{\mathrm{TX}}(f)$ is multiplied with the channel function $H_{12}(f)$ to get $S_{\mathrm{RX}}(f)=H_{12}(f) S_{\mathrm{TX}}(f)$.
The received signal in time $s_{\mathrm{RX}}(t)$ can be calculated as the real part of the inverse Fourier transform of $S_{\mathrm{RX}}(f)$. By using this procedure, $s_{\text{TX}}(t)$ is convoluted with $h_{12}(t)$, the channel impulse response. It essentially represents how the channel affects the signal over time, including effects like multipath propagation, fading, and delay.
Furthermore, the Gaussian noise $n(t)$ of zero mean and standard deviation $\chi$ is added to the signal on the receiver side. It can emulate thermal noise at the receiver side independent of the received signal and the noise power will be $P_{\mathrm{n}}=10 \log _{10}\left(\chi^2 / 10^{-3}\right) \mathrm{dBm}$. It is therefore possible to express the received signal as follows:

\begin{equation}
	s_{\mathrm{RX}}(t)=s_{\mathrm{TX}}(t) * h_{12}(t)+n(t).
\end{equation}

Thirdly, the received signal $s_\mathrm{RX}(t)$ undergoes sampling and quantization to produce the discrete signal stream $\mathbf{b}_\mathrm{r}$. After this step, $s_\mathrm{RX}(t)$ is segmented into intervals of duration $\tau_\text{s}$, which are shifted relative to the transmitted signal by a propagation delay $\Delta_{\mathrm{pd}}$, corresponding to the time delay in detecting the first received pulse.
Subsequently, energy detection is performed by integrating $s_\mathrm{RX}^2(t)$ over each time interval of $\tau_\text{s}$, resulting in a $1 \times N$ vector representing the detected energy,

\begin{equation}
	\mathbf{d}[i]=\int_{(i-1) \tau_\text{s}+\Delta_{\mathrm{pd}}}^{i \tau_\text{s}+\Delta_{\mathrm{pd}}} s_{\mathrm{RX}}^2(t)\:\text{d}t.
\end{equation}
After this, each value of $\mathbf{d}$ is evaluated by comparison to a threshold value $d_\text{th}$, and "1" is recorded if the value has exceeded the threshold, and "0" otherwise,
\begin{equation}
\mathbf{b}_{\mathbf{r}}[i]=\left\{ \begin{array}{l}
0\text{, if }\mathrm{d}[i]\leq d_{\text{th }}\\
1\text{, if }\mathrm{d}[i]>d_{\text{th }}
\end{array}\right.
\end{equation}
This results in the decoded sequence of $b_\mathrm{r}$. Over a long series of random symbols, the threshold is determined as the energy value that minimizes the BER value.


\section{Multi-channel Metasurface for Advanced Information Encryption}
The proposed metasurface features can be used for more powerful operations such as electromagnetic information transmission in multi-frequency channels and power controls by arranging meta-atoms on the metasurface. We examine scattering patterns across two frequency channels employing distinct coding sequences. As shown in Fig. \ref{fig: Scattering}, employing a $20 \times 20$ meta-atom arrangement and individual control over elements via the FPGA, we can exhibit various scattering pattern states using the reprogrammable amplitude coding metasurface with multi-frequency modulation. The metasurface arrangement utilizes 2-bit meta-atoms, illustrated in Fig. \ref{fig: UC}(b), leading to four unique states for the scattering patterns. 
In Fig. \ref{fig: Scattering}, the scattering patterns shown in red and blue correspond to far-field outcomes observed at frequencies $f_1$ and $f_1$, respectively. Also, simulation results were obtained using CST Studio Suite. To investigate different scattering patterns, we encode the metasurface with elements "$\text{L}0/\text{H}0$", "$\text{L}0/\text{H}1$", "$\text{L}1/\text{H}0$", and "$\text{L}1/\text{H}/1$" separately. As shown in Fig. \ref{fig: Scattering}(a), because all the elements of the metasurface array are in the "$\text{L}0/\text{H}0$" state, the incident wave is absorbed at both $f_1$ and $f_2$ frequencies. Consequently, the amplitude profile of the scattering pattern is notably low at both frequencies. Similarly, in Figs. \ref{fig: Scattering}(b)–(d), all element of the metasurface array is individually set to the "$\text{L}0/\text{H}1$", "$\text{L}1/\text{H}0$", and "$\text{L}1/\text{H}1$" states, resulting in varying amplitude profiles on the different scattering patterns. By adjusting the metasurface amplitude in the OOK format across two different frequency channels, we can transmit our information independently through each channel or simultaneously through both channels. To safeguard against information leakage by eavesdroppers, we employ an encryption method to secure the transmitted information, ensuring that even if intercepted, it remains undisclosed information without the corresponding decryption key.

\begin{figure*}
	\centering
	\includegraphics[width=0.95\textwidth]{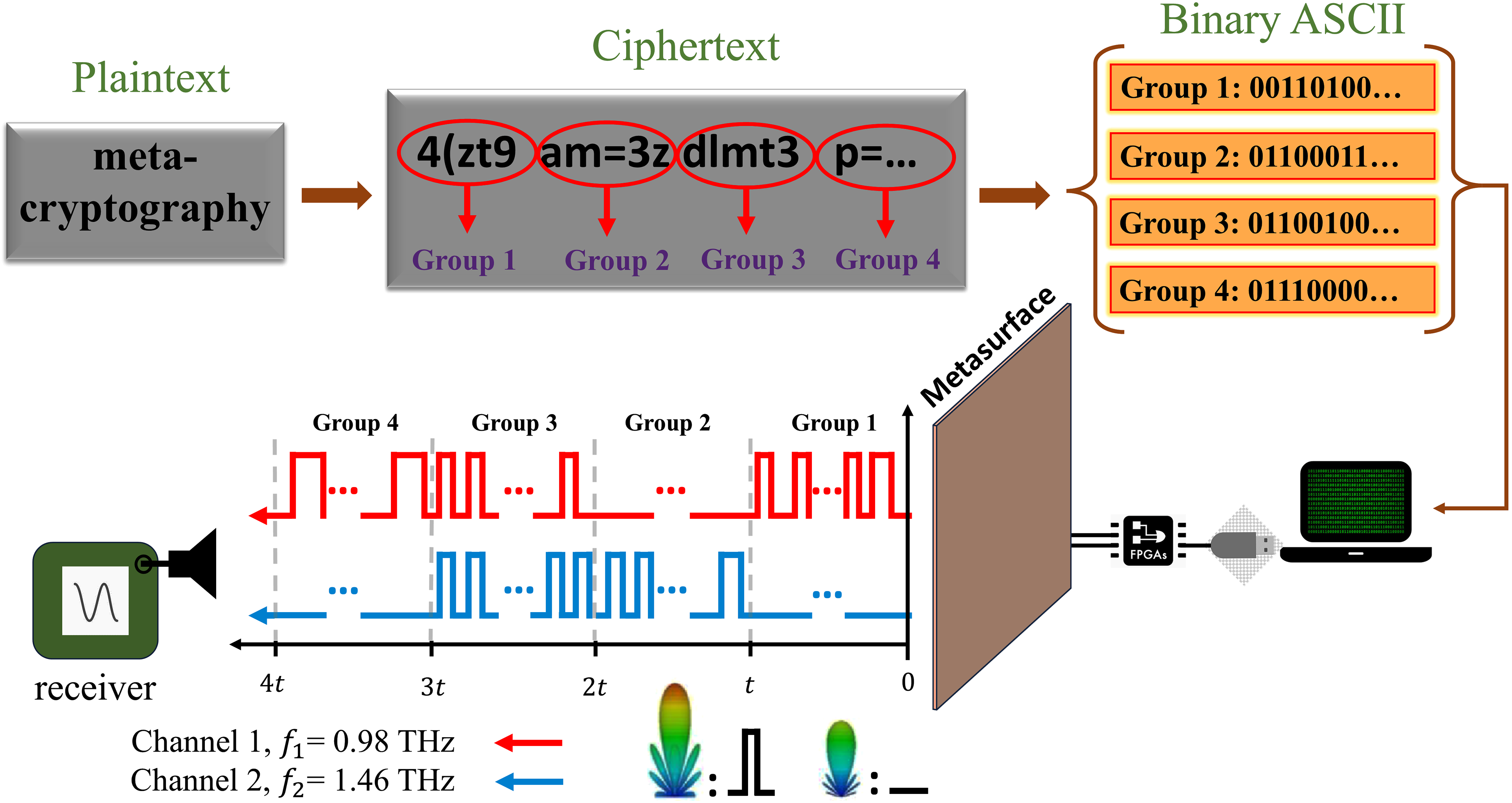}
	\caption{The purpose information, meta-cryptography, is encrypted by substitution cipher method with alphabets/symbols which produce four groups of five. Each alphabet/symbol converts to its own corresponding binary ASCII code. Eventually, produced bits are transmitted through FPGA to the metasurface to control amplitude-frequency modulation in real-time for transmitting each group of target information at a specific time period.}
	\label{fig: Crypto}
\end{figure*}

We can encrypt our information using a substitution cipher method with symbols, which involves replacing single letters with different letters or symbols \cite{14,15}. The key to this method is a mapping that connects letters to other symbols or letters. This key is transmitted through a private channel between the transmitter and receiver, as shown schematically in Fig. \ref{fig: Main}. As depicted in Table \ref{tab: Cipher}, each letter of the alphabet, along with numbers and symbols, is associated with another letter, number, or symbol. This process encrypts the information before it is transmitted by the sender. For instance, when transmitting the phrase "Parsa and Kasra!", we encrypt it using a predefined dictionary. Consequently, instead of sending the original phrase, we send the ciphertext "3tm?t t2s et?mti". Even if an eavesdropper intercepts the transmitted information, without the decryption key, they remain unable to decipher it, thus failing to comprehend the intended message. It's important to note that encrypted text is often organized into fixed-length groups, with punctuation marks and spaces commonly omitted. Also, we do not distinguish between capital and small letters in our dictionary. This practice hides word boundaries in plaintext, adding complexity to the encrypted data and helping prevent transmission errors. The length of these groups can vary, but historically, when information was transmitted via telegraph, these groups were typically five letters long. Therefore, we adhere to using encrypted information groups of five letters in length. It's important to remember that if the message length is not divisible by five, one or more "." symbols can be added at the end of the group to ensure it consists of five symbols. In this case, the receiver can readily identify the additional "." symbols and remove them.

Using the reprogrammable amplitude-coding metasurface, we propose to design an OOK communication demonstration as a proof of concept. As depicted in Fig. \ref{fig: Crypto}, the term "meta-cryptography" undergoes encryption by the substitution cipher method, resulting in 17 ciphertext characters. To ensure our ciphertext is organized into groups of five phrases, we add three "." phrases to the end of the ciphertext. This adjustment results in four groups, each containing five phrases. As information is transmitted in bits using amplitude encryption, it is essential to convert ciphertext into binary bits. To achieve this, we replace each phrase with its corresponding binary ASCII code. Consequently, each phrase contains 8 bits of information, resulting in a total of 40 bits for each group. To increase the complexity of the encrypted information for potential eavesdroppers, we send the information across two frequency channels, $f_1$ and $f_2$, in three distinct ways: (1) sending information solely in the $f_1$ frequency channel, leaving the $f_2$ frequency channel devoid of transmitted information; (2) sending information exclusively in the $f_2$ frequency channel, leaving the $f_1$ frequency channel empty of transmitted information; and (3) simultaneously transmitting information across both $f_1$ and $f_2$ frequencies. To transmit information with amplitude control as "0" and "1" exclusively on the $f_1$ frequency, while no information is sent on the $f_2$ frequency, we need to utilize the $\text{L}0/\text{H}0$ and $\text{L}1/\text{H}0$ coding states. Here, the amplitude is modulated in binary form at frequency $f_1$, while absorption occurs at frequency $f_2$ in both states. Consequently, no information is transmitted at frequency $f_2$. Conversely, in the $\text{L}0/\text{H}0$ and $\text{L}0/\text{H}1$ coding states, the amplitude is modulated in binary form at the $f_2$ frequency, while at the $f_1$ frequency, it is absorbed in both states. This indicates that no information is transmitted at the $f_1$ frequency in these modes. For the mode of transmitting information simultaneously at both $f_1$ and $f_2$ frequencies, we employ the same coding state depicted in Fig. \ref{fig: UC}(b). 

In the first two methods of communicating information, the amplitude is controlled by only 1-bit at a single frequency. However, when transmitting information simultaneously across both frequencies, the amplitude is controlled by 2-bit. Given that we can transmit information across two distinct frequency channels in three different ways, each ciphertext group is transmitted using one of these methods. To establish a structured transmission of encrypted groups, an agreement is reached between the sender and receiver. This agreement dictates that the first group is transmitted via the $f_1$ frequency channel, the second group via the $f_2$ frequency channel, and the third group via both the $f_1$ and $f_2$ frequency channels simultaneously. This sequential order is maintained for subsequent groups as well. To transmit information, the binary bits derived from the binary ASCII codes of the encrypted data are conveyed to the metasurface through the FPGA. Subsequently, the proposed metasurface transmits information in accordance with the provided instructions. It is important to note that sending the information of each group in each frequency channel takes $t_d$ seconds. Consequently, the receiver detects the channel change based on the specified time interval after its completion and avoids misinterpreting the zero bit sent in the preceding frequency channel after $t_d$ seconds. On the receiver side, the binary waveform sent in both frequency channels can be demultiplexed, allowing for the recording of the transmitted binary bits. In this scenario, the receiver can decrypt the received bits using the key already sent via the private channel from the sender.

\section{conclusion}
We presented a reprogrammable metasurface that utilizes OOK modulation with independent amplitude and frequency control at two THz frequencies. The metasurface, composed of two graphene layers with separate DC biases, enables dynamic 1-bit and 2-bit control of amplitude and frequency through an FPGA. To ensure secure communication, we applied the substitution cipher method to encrypt data transmitted at distinct frequencies, $f_1 = 0.98\:\text{THz}$ and $f_2 = 1.46\:\text{THz}$. The simulation results demonstrate that this programmable metasurface can enhances information security while offering flexibility for multi-channel communication, encryption, and advanced data processing in the THz domain


\begin{thebibliography}{72}%
	\makeatletter
	\providecommand \@ifxundefined [1]{%
		\@ifx{#1\undefined}
	}%
	\providecommand \@ifnum [1]{%
		\ifnum #1\expandafter \@firstoftwo
		\else \expandafter \@secondoftwo
		\fi
	}%
	\providecommand \@ifx [1]{%
		\ifx #1\expandafter \@firstoftwo
		\else \expandafter \@secondoftwo
		\fi
	}%
	\providecommand \natexlab [1]{#1}%
	\providecommand \enquote  [1]{``#1''}%
	\providecommand \bibnamefont  [1]{#1}%
	\providecommand \bibfnamefont [1]{#1}%
	\providecommand \citenamefont [1]{#1}%
	\providecommand \href@noop [0]{\@secondoftwo}%
	\providecommand \href [0]{\begingroup \@sanitize@url \@href}%
	\providecommand \@href[1]{\@@startlink{#1}\@@href}%
	\providecommand \@@href[1]{\endgroup#1\@@endlink}%
	\providecommand \@sanitize@url [0]{\catcode `\\12\catcode `\$12\catcode
		`\&12\catcode `\#12\catcode `\^12\catcode `\_12\catcode `\%12\relax}%
	\providecommand \@@startlink[1]{}%
	\providecommand \@@endlink[0]{}%
	\providecommand \url  [0]{\begingroup\@sanitize@url \@url }%
	\providecommand \@url [1]{\endgroup\@href {#1}{\urlprefix }}%
	\providecommand \urlprefix  [0]{URL }%
	\providecommand \Eprint [0]{\href }%
	\providecommand \doibase [0]{https://doi.org/}%
	\providecommand \selectlanguage [0]{\@gobble}%
	\providecommand \bibinfo  [0]{\@secondoftwo}%
	\providecommand \bibfield  [0]{\@secondoftwo}%
	\providecommand \translation [1]{[#1]}%
	\providecommand \BibitemOpen [0]{}%
	\providecommand \bibitemStop [0]{}%
	\providecommand \bibitemNoStop [0]{.\EOS\space}%
	\providecommand \EOS [0]{\spacefactor3000\relax}%
	\providecommand \BibitemShut  [1]{\csname bibitem#1\endcsname}%
	\let\auto@bib@innerbib\@empty
	\bibitem [{\citenamefont {Kleine-Ostmann}\ and\ \citenamefont
		{Nagatsuma}(2011)}]{kleine2011review}%
	\BibitemOpen
	\bibfield  {author} {\bibinfo {author} {\bibfnamefont {T.}~\bibnamefont
			{Kleine-Ostmann}}\ and\ \bibinfo {author} {\bibfnamefont {T.}~\bibnamefont
			{Nagatsuma}},\ }\bibfield  {title} {\bibinfo {title} {A review on terahertz
			communications research},\ }\href@noop {} {\bibfield  {journal} {\bibinfo
			{journal} {Journal of Infrared, Millimeter, and Terahertz Waves}\ }\textbf
		{\bibinfo {volume} {32}},\ \bibinfo {pages} {143} (\bibinfo {year}
		{2011})}\BibitemShut {NoStop}%
	\bibitem [{\citenamefont {Lewis}(2014)}]{lewis2014review}%
	\BibitemOpen
	\bibfield  {author} {\bibinfo {author} {\bibfnamefont {R.~A.}\ \bibnamefont
			{Lewis}},\ }\bibfield  {title} {\bibinfo {title} {A review of terahertz
			sources},\ }\href@noop {} {\bibfield  {journal} {\bibinfo  {journal} {Journal
				of Physics D: Applied Physics}\ }\textbf {\bibinfo {volume} {47}},\ \bibinfo
		{pages} {374001} (\bibinfo {year} {2014})}\BibitemShut {NoStop}%
	\bibitem [{\citenamefont {Monticone}\ and\ \citenamefont
		{Alu}(2017)}]{monticone2017metamaterial}%
	\BibitemOpen
	\bibfield  {author} {\bibinfo {author} {\bibfnamefont {F.}~\bibnamefont
			{Monticone}}\ and\ \bibinfo {author} {\bibfnamefont {A.}~\bibnamefont
			{Alu}},\ }\bibfield  {title} {\bibinfo {title} {Metamaterial, plasmonic and
			nanophotonic devices},\ }\href@noop {} {\bibfield  {journal} {\bibinfo
			{journal} {Reports on Progress in Physics}\ }\textbf {\bibinfo {volume}
			{80}},\ \bibinfo {pages} {036401} (\bibinfo {year} {2017})}\BibitemShut
	{NoStop}%
	\bibitem [{\citenamefont {Xu}\ \emph {et~al.}(2021)\citenamefont {Xu},
		\citenamefont {Pu}, \citenamefont {Sang}, \citenamefont {Zheng},
		\citenamefont {Li}, \citenamefont {Ma}, \citenamefont {Guo}, \citenamefont
		{Zhang},\ and\ \citenamefont {Luo}}]{26}%
	\BibitemOpen
	\bibfield  {author} {\bibinfo {author} {\bibfnamefont {M.}~\bibnamefont
			{Xu}}, \bibinfo {author} {\bibfnamefont {M.}~\bibnamefont {Pu}}, \bibinfo
		{author} {\bibfnamefont {D.}~\bibnamefont {Sang}}, \bibinfo {author}
		{\bibfnamefont {Y.}~\bibnamefont {Zheng}}, \bibinfo {author} {\bibfnamefont
			{X.}~\bibnamefont {Li}}, \bibinfo {author} {\bibfnamefont {X.}~\bibnamefont
			{Ma}}, \bibinfo {author} {\bibfnamefont {Y.}~\bibnamefont {Guo}}, \bibinfo
		{author} {\bibfnamefont {R.}~\bibnamefont {Zhang}},\ and\ \bibinfo {author}
		{\bibfnamefont {X.}~\bibnamefont {Luo}},\ }\bibfield  {title} {\bibinfo
		{title} {Topology-optimized catenary-like metasurface for wide-angle and
			high-efficiency deflection: from a discrete to continuous geometric phase},\
	}\href@noop {} {\bibfield  {journal} {\bibinfo  {journal} {Optics Express}\
		}\textbf {\bibinfo {volume} {29}},\ \bibinfo {pages} {10181} (\bibinfo {year}
		{2021})}\BibitemShut {NoStop}%
	\bibitem [{\citenamefont {Chen}\ \emph {et~al.}(2016)\citenamefont {Chen},
		\citenamefont {Taylor},\ and\ \citenamefont {Yu}}]{27}%
	\BibitemOpen
	\bibfield  {author} {\bibinfo {author} {\bibfnamefont {H.-T.}\ \bibnamefont
			{Chen}}, \bibinfo {author} {\bibfnamefont {A.~J.}\ \bibnamefont {Taylor}},\
		and\ \bibinfo {author} {\bibfnamefont {N.}~\bibnamefont {Yu}},\ }\bibfield
	{title} {\bibinfo {title} {A review of metasurfaces: physics and
			applications},\ }\href@noop {} {\bibfield  {journal} {\bibinfo  {journal}
			{Reports on progress in physics}\ }\textbf {\bibinfo {volume} {79}},\
		\bibinfo {pages} {076401} (\bibinfo {year} {2016})}\BibitemShut {NoStop}%
	\bibitem [{\citenamefont {Glybovski}\ \emph {et~al.}(2016)\citenamefont
		{Glybovski}, \citenamefont {Tretyakov}, \citenamefont {Belov}, \citenamefont
		{Kivshar},\ and\ \citenamefont {Simovski}}]{28}%
	\BibitemOpen
	\bibfield  {author} {\bibinfo {author} {\bibfnamefont {S.~B.}\ \bibnamefont
			{Glybovski}}, \bibinfo {author} {\bibfnamefont {S.~A.}\ \bibnamefont
			{Tretyakov}}, \bibinfo {author} {\bibfnamefont {P.~A.}\ \bibnamefont
			{Belov}}, \bibinfo {author} {\bibfnamefont {Y.~S.}\ \bibnamefont {Kivshar}},\
		and\ \bibinfo {author} {\bibfnamefont {C.~R.}\ \bibnamefont {Simovski}},\
	}\bibfield  {title} {\bibinfo {title} {Metasurfaces: From microwaves to
			visible},\ }\href@noop {} {\bibfield  {journal} {\bibinfo  {journal} {Physics
				reports}\ }\textbf {\bibinfo {volume} {634}},\ \bibinfo {pages} {1} (\bibinfo
		{year} {2016})}\BibitemShut {NoStop}%
	\bibitem [{\citenamefont {Holloway}\ \emph {et~al.}(2012)\citenamefont
		{Holloway}, \citenamefont {Kuester}, \citenamefont {Gordon}, \citenamefont
		{O'Hara}, \citenamefont {Booth},\ and\ \citenamefont {Smith}}]{29}%
	\BibitemOpen
	\bibfield  {author} {\bibinfo {author} {\bibfnamefont {C.~L.}\ \bibnamefont
			{Holloway}}, \bibinfo {author} {\bibfnamefont {E.~F.}\ \bibnamefont
			{Kuester}}, \bibinfo {author} {\bibfnamefont {J.~A.}\ \bibnamefont {Gordon}},
		\bibinfo {author} {\bibfnamefont {J.}~\bibnamefont {O'Hara}}, \bibinfo
		{author} {\bibfnamefont {J.}~\bibnamefont {Booth}},\ and\ \bibinfo {author}
		{\bibfnamefont {D.~R.}\ \bibnamefont {Smith}},\ }\bibfield  {title} {\bibinfo
		{title} {An overview of the theory and applications of metasurfaces: The
			two-dimensional equivalents of metamaterials},\ }\href@noop {} {\bibfield
		{journal} {\bibinfo  {journal} {IEEE antennas and propagation magazine}\
		}\textbf {\bibinfo {volume} {54}},\ \bibinfo {pages} {10} (\bibinfo {year}
		{2012})}\BibitemShut {NoStop}%
	\bibitem [{\citenamefont {Hu}\ \emph {et~al.}(2021)\citenamefont {Hu},
		\citenamefont {Bandyopadhyay}, \citenamefont {Liu},\ and\ \citenamefont
		{Shao}}]{hu2021review}%
	\BibitemOpen
	\bibfield  {author} {\bibinfo {author} {\bibfnamefont {J.}~\bibnamefont
			{Hu}}, \bibinfo {author} {\bibfnamefont {S.}~\bibnamefont {Bandyopadhyay}},
		\bibinfo {author} {\bibfnamefont {Y.-h.}\ \bibnamefont {Liu}},\ and\ \bibinfo
		{author} {\bibfnamefont {L.-y.}\ \bibnamefont {Shao}},\ }\bibfield  {title}
	{\bibinfo {title} {A review on metasurface: from principle to smart
			metadevices},\ }\href@noop {} {\bibfield  {journal} {\bibinfo  {journal}
			{Frontiers in Physics}\ }\textbf {\bibinfo {volume} {8}},\ \bibinfo {pages}
		{586087} (\bibinfo {year} {2021})}\BibitemShut {NoStop}%
	\bibitem [{\citenamefont {Hosseininejad}\ \emph {et~al.}(2019)\citenamefont
		{Hosseininejad}, \citenamefont {Rouhi}, \citenamefont {Neshat}, \citenamefont
		{Cabellos-Aparicio}, \citenamefont {Abadal},\ and\ \citenamefont
		{Alarc{\'o}n}}]{33}%
	\BibitemOpen
	\bibfield  {author} {\bibinfo {author} {\bibfnamefont {S.~E.}\ \bibnamefont
			{Hosseininejad}}, \bibinfo {author} {\bibfnamefont {K.}~\bibnamefont
			{Rouhi}}, \bibinfo {author} {\bibfnamefont {M.}~\bibnamefont {Neshat}},
		\bibinfo {author} {\bibfnamefont {A.}~\bibnamefont {Cabellos-Aparicio}},
		\bibinfo {author} {\bibfnamefont {S.}~\bibnamefont {Abadal}},\ and\ \bibinfo
		{author} {\bibfnamefont {E.}~\bibnamefont {Alarc{\'o}n}},\ }\bibfield
	{title} {\bibinfo {title} {Digital metasurface based on graphene: An
			application to beam steering in terahertz plasmonic antennas},\ }\href@noop
	{} {\bibfield  {journal} {\bibinfo  {journal} {IEEE Transactions on
				Nanotechnology}\ }\textbf {\bibinfo {volume} {18}},\ \bibinfo {pages} {734}
		(\bibinfo {year} {2019})}\BibitemShut {NoStop}%
	\bibitem [{\citenamefont {Sun}\ \emph {et~al.}(2012)\citenamefont {Sun},
		\citenamefont {He}, \citenamefont {Xiao}, \citenamefont {Xu}, \citenamefont
		{Li},\ and\ \citenamefont {Zhou}}]{34}%
	\BibitemOpen
	\bibfield  {author} {\bibinfo {author} {\bibfnamefont {S.}~\bibnamefont
			{Sun}}, \bibinfo {author} {\bibfnamefont {Q.}~\bibnamefont {He}}, \bibinfo
		{author} {\bibfnamefont {S.}~\bibnamefont {Xiao}}, \bibinfo {author}
		{\bibfnamefont {Q.}~\bibnamefont {Xu}}, \bibinfo {author} {\bibfnamefont
			{X.}~\bibnamefont {Li}},\ and\ \bibinfo {author} {\bibfnamefont
			{L.}~\bibnamefont {Zhou}},\ }\bibfield  {title} {\bibinfo {title}
		{Gradient-index meta-surfaces as a bridge linking propagating waves and
			surface waves},\ }\href@noop {} {\bibfield  {journal} {\bibinfo  {journal}
			{Nature materials}\ }\textbf {\bibinfo {volume} {11}},\ \bibinfo {pages}
		{426} (\bibinfo {year} {2012})}\BibitemShut {NoStop}%
	\bibitem [{\citenamefont {Wang}\ \emph {et~al.}(2018)\citenamefont {Wang},
		\citenamefont {Wu}, \citenamefont {Su}, \citenamefont {Lai}, \citenamefont
		{Chen}, \citenamefont {Kuo}, \citenamefont {Chen}, \citenamefont {Chen},
		\citenamefont {Huang}, \citenamefont {Wang} \emph {et~al.}}]{35}%
	\BibitemOpen
	\bibfield  {author} {\bibinfo {author} {\bibfnamefont {S.}~\bibnamefont
			{Wang}}, \bibinfo {author} {\bibfnamefont {P.~C.}\ \bibnamefont {Wu}},
		\bibinfo {author} {\bibfnamefont {V.-C.}\ \bibnamefont {Su}}, \bibinfo
		{author} {\bibfnamefont {Y.-C.}\ \bibnamefont {Lai}}, \bibinfo {author}
		{\bibfnamefont {M.-K.}\ \bibnamefont {Chen}}, \bibinfo {author}
		{\bibfnamefont {H.~Y.}\ \bibnamefont {Kuo}}, \bibinfo {author} {\bibfnamefont
			{B.~H.}\ \bibnamefont {Chen}}, \bibinfo {author} {\bibfnamefont {Y.~H.}\
			\bibnamefont {Chen}}, \bibinfo {author} {\bibfnamefont {T.-T.}\ \bibnamefont
			{Huang}}, \bibinfo {author} {\bibfnamefont {J.-H.}\ \bibnamefont {Wang}},
		\emph {et~al.},\ }\bibfield  {title} {\bibinfo {title} {A broadband
			achromatic metalens in the visible},\ }\href@noop {} {\bibfield  {journal}
		{\bibinfo  {journal} {Nature nanotechnology}\ }\textbf {\bibinfo {volume}
			{13}},\ \bibinfo {pages} {227} (\bibinfo {year} {2018})}\BibitemShut
	{NoStop}%
	\bibitem [{\citenamefont {Akram}\ \emph {et~al.}(2020)\citenamefont {Akram},
		\citenamefont {Ding}, \citenamefont {Chen}, \citenamefont {Feng},\ and\
		\citenamefont {Zhu}}]{36}%
	\BibitemOpen
	\bibfield  {author} {\bibinfo {author} {\bibfnamefont {M.~R.}\ \bibnamefont
			{Akram}}, \bibinfo {author} {\bibfnamefont {G.}~\bibnamefont {Ding}},
		\bibinfo {author} {\bibfnamefont {K.}~\bibnamefont {Chen}}, \bibinfo {author}
		{\bibfnamefont {Y.}~\bibnamefont {Feng}},\ and\ \bibinfo {author}
		{\bibfnamefont {W.}~\bibnamefont {Zhu}},\ }\bibfield  {title} {\bibinfo
		{title} {Ultrathin single layer metasurfaces with ultra-wideband operation
			for both transmission and reflection},\ }\href@noop {} {\bibfield  {journal}
		{\bibinfo  {journal} {Advanced Materials}\ }\textbf {\bibinfo {volume}
			{32}},\ \bibinfo {pages} {1907308} (\bibinfo {year} {2020})}\BibitemShut
	{NoStop}%
	\bibitem [{\citenamefont {Lee}\ and\ \citenamefont {Kwon}(2022)}]{37}%
	\BibitemOpen
	\bibfield  {author} {\bibinfo {author} {\bibfnamefont {H.}~\bibnamefont
			{Lee}}\ and\ \bibinfo {author} {\bibfnamefont {D.-H.}\ \bibnamefont {Kwon}},\
	}\bibfield  {title} {\bibinfo {title} {Microwave metasurface cloaking for
			freestanding objects},\ }\href@noop {} {\bibfield  {journal} {\bibinfo
			{journal} {Physical Review Applied}\ }\textbf {\bibinfo {volume} {17}},\
		\bibinfo {pages} {054012} (\bibinfo {year} {2022})}\BibitemShut {NoStop}%
	\bibitem [{\citenamefont {Jahangiri}\ \emph {et~al.}(2024)\citenamefont
		{Jahangiri}, \citenamefont {Soleimani},\ and\ \citenamefont
		{Soleimani}}]{jahangiri2024beam}%
	\BibitemOpen
	\bibfield  {author} {\bibinfo {author} {\bibfnamefont {M.}~\bibnamefont
			{Jahangiri}}, \bibinfo {author} {\bibfnamefont {H.}~\bibnamefont
			{Soleimani}},\ and\ \bibinfo {author} {\bibfnamefont {M.}~\bibnamefont
			{Soleimani}},\ }\bibfield  {title} {\bibinfo {title} {Beam steering by two
			vertically faced metasurfaces using polarization-free unit cells with three
			operating modes},\ }\href@noop {} {\bibfield  {journal} {\bibinfo  {journal}
			{arXiv preprint arXiv:2405.05976}\ } (\bibinfo {year} {2024})}\BibitemShut
	{NoStop}%
	\bibitem [{\citenamefont {Farzin}\ \emph
		{et~al.}(2024{\natexlab{a}})\citenamefont {Farzin}, \citenamefont
		{Nooramin},\ and\ \citenamefont {Soleimani}}]{farzin2024reprogrammable}%
	\BibitemOpen
	\bibfield  {author} {\bibinfo {author} {\bibfnamefont {P.}~\bibnamefont
			{Farzin}}, \bibinfo {author} {\bibfnamefont {A.~S.}\ \bibnamefont
			{Nooramin}},\ and\ \bibinfo {author} {\bibfnamefont {M.}~\bibnamefont
			{Soleimani}},\ }\bibfield  {title} {\bibinfo {title} {Reprogrammable
			reflection-transmission integrated coding metasurface for real-time terahertz
			wavefront manipulations in full-space},\ }\href@noop {} {\bibfield  {journal}
		{\bibinfo  {journal} {Scientific Reports}\ }\textbf {\bibinfo {volume}
			{14}},\ \bibinfo {pages} {11156} (\bibinfo {year}
		{2024}{\natexlab{a}})}\BibitemShut {NoStop}%
	\bibitem [{\citenamefont {Ni}\ \emph {et~al.}(2013)\citenamefont {Ni},
		\citenamefont {Kildishev},\ and\ \citenamefont {Shalaev}}]{38}%
	\BibitemOpen
	\bibfield  {author} {\bibinfo {author} {\bibfnamefont {X.}~\bibnamefont
			{Ni}}, \bibinfo {author} {\bibfnamefont {A.~V.}\ \bibnamefont {Kildishev}},\
		and\ \bibinfo {author} {\bibfnamefont {V.~M.}\ \bibnamefont {Shalaev}},\
	}\bibfield  {title} {\bibinfo {title} {Metasurface holograms for visible
			light},\ }\href@noop {} {\bibfield  {journal} {\bibinfo  {journal} {Nature
				communications}\ }\textbf {\bibinfo {volume} {4}},\ \bibinfo {pages} {2807}
		(\bibinfo {year} {2013})}\BibitemShut {NoStop}%
	\bibitem [{\citenamefont {Cui}\ \emph {et~al.}(2014)\citenamefont {Cui},
		\citenamefont {Qi}, \citenamefont {Wan}, \citenamefont {Zhao},\ and\
		\citenamefont {Cheng}}]{cui2014coding}%
	\BibitemOpen
	\bibfield  {author} {\bibinfo {author} {\bibfnamefont {T.~J.}\ \bibnamefont
			{Cui}}, \bibinfo {author} {\bibfnamefont {M.~Q.}\ \bibnamefont {Qi}},
		\bibinfo {author} {\bibfnamefont {X.}~\bibnamefont {Wan}}, \bibinfo {author}
		{\bibfnamefont {J.}~\bibnamefont {Zhao}},\ and\ \bibinfo {author}
		{\bibfnamefont {Q.}~\bibnamefont {Cheng}},\ }\bibfield  {title} {\bibinfo
		{title} {Coding metamaterials, digital metamaterials and programmable
			metamaterials},\ }\href@noop {} {\bibfield  {journal} {\bibinfo  {journal}
			{Light: science \& applications}\ }\textbf {\bibinfo {volume} {3}},\ \bibinfo
		{pages} {e218} (\bibinfo {year} {2014})}\BibitemShut {NoStop}%
	\bibitem [{\citenamefont {Tan}\ \emph {et~al.}(2022)\citenamefont {Tan},
		\citenamefont {Chen}, \citenamefont {Li},\ and\ \citenamefont {Yan}}]{40}%
	\BibitemOpen
	\bibfield  {author} {\bibinfo {author} {\bibfnamefont {X.}~\bibnamefont
			{Tan}}, \bibinfo {author} {\bibfnamefont {J.}~\bibnamefont {Chen}}, \bibinfo
		{author} {\bibfnamefont {J.}~\bibnamefont {Li}},\ and\ \bibinfo {author}
		{\bibfnamefont {S.}~\bibnamefont {Yan}},\ }\bibfield  {title} {\bibinfo
		{title} {Water-based metasurface with continuously tunable reflection
			amplitude},\ }\href@noop {} {\bibfield  {journal} {\bibinfo  {journal}
			{Optics Express}\ }\textbf {\bibinfo {volume} {30}},\ \bibinfo {pages} {6991}
		(\bibinfo {year} {2022})}\BibitemShut {NoStop}%
	\bibitem [{\citenamefont {Momeni}\ \emph {et~al.}(2018)\citenamefont {Momeni},
		\citenamefont {Rouhi}, \citenamefont {Rajabalipanah},\ and\ \citenamefont
		{Abdolali}}]{41}%
	\BibitemOpen
	\bibfield  {author} {\bibinfo {author} {\bibfnamefont {A.}~\bibnamefont
			{Momeni}}, \bibinfo {author} {\bibfnamefont {K.}~\bibnamefont {Rouhi}},
		\bibinfo {author} {\bibfnamefont {H.}~\bibnamefont {Rajabalipanah}},\ and\
		\bibinfo {author} {\bibfnamefont {A.}~\bibnamefont {Abdolali}},\ }\bibfield
	{title} {\bibinfo {title} {An information theory-inspired strategy for design
			of re-programmable encrypted graphene-based coding metasurfaces at terahertz
			frequencies},\ }\href@noop {} {\bibfield  {journal} {\bibinfo  {journal}
			{Scientific reports}\ }\textbf {\bibinfo {volume} {8}},\ \bibinfo {pages} {1}
		(\bibinfo {year} {2018})}\BibitemShut {NoStop}%
	\bibitem [{\citenamefont {Farzin}\ \emph
		{et~al.}(2024{\natexlab{b}})\citenamefont {Farzin}, \citenamefont
		{Hajiahmadi},\ and\ \citenamefont {Soleimani}}]{farzin2024multi}%
	\BibitemOpen
	\bibfield  {author} {\bibinfo {author} {\bibfnamefont {P.}~\bibnamefont
			{Farzin}}, \bibinfo {author} {\bibfnamefont {M.~J.}\ \bibnamefont
			{Hajiahmadi}},\ and\ \bibinfo {author} {\bibfnamefont {M.}~\bibnamefont
			{Soleimani}},\ }\bibfield  {title} {\bibinfo {title} {Multi-channel
			polarization manipulation based on graphene for encryption communication},\
	}\href@noop {} {\bibfield  {journal} {\bibinfo  {journal} {Scientific
				Reports}\ }\textbf {\bibinfo {volume} {14}},\ \bibinfo {pages} {11155}
		(\bibinfo {year} {2024}{\natexlab{b}})}\BibitemShut {NoStop}%
	\bibitem [{\citenamefont {Ozgoli}\ \emph {et~al.}(2024)\citenamefont {Ozgoli},
		\citenamefont {Farzin}, \citenamefont {Hajiahmadi},\ and\ \citenamefont
		{Soleimani}}]{ozgoli2024multichannel}%
	\BibitemOpen
	\bibfield  {author} {\bibinfo {author} {\bibfnamefont {A.}~\bibnamefont
			{Ozgoli}}, \bibinfo {author} {\bibfnamefont {P.}~\bibnamefont {Farzin}},
		\bibinfo {author} {\bibfnamefont {M.~J.}\ \bibnamefont {Hajiahmadi}},\ and\
		\bibinfo {author} {\bibfnamefont {M.}~\bibnamefont {Soleimani}},\ }\bibfield
	{title} {\bibinfo {title} {Multichannel
			joint-polarization-frequency-modulation encrypted metasurface in secure thz
			communication},\ }\href@noop {} {\bibfield  {journal} {\bibinfo  {journal}
			{arXiv preprint arXiv:2409.15966}\ } (\bibinfo {year} {2024})}\BibitemShut
	{NoStop}%
	\bibitem [{\citenamefont {Zheng}\ \emph {et~al.}(2022)\citenamefont {Zheng},
		\citenamefont {Chen}, \citenamefont {Xu}, \citenamefont {Zhang},
		\citenamefont {Wang}, \citenamefont {Zhao},\ and\ \citenamefont {Feng}}]{44}%
	\BibitemOpen
	\bibfield  {author} {\bibinfo {author} {\bibfnamefont {Y.}~\bibnamefont
			{Zheng}}, \bibinfo {author} {\bibfnamefont {K.}~\bibnamefont {Chen}},
		\bibinfo {author} {\bibfnamefont {Z.}~\bibnamefont {Xu}}, \bibinfo {author}
		{\bibfnamefont {N.}~\bibnamefont {Zhang}}, \bibinfo {author} {\bibfnamefont
			{J.}~\bibnamefont {Wang}}, \bibinfo {author} {\bibfnamefont {J.}~\bibnamefont
			{Zhao}},\ and\ \bibinfo {author} {\bibfnamefont {Y.}~\bibnamefont {Feng}},\
	}\bibfield  {title} {\bibinfo {title} {Metasurface-assisted wireless
			communication with physical level information encryption},\ }\href@noop {}
	{\bibfield  {journal} {\bibinfo  {journal} {Advanced Science}\ }\textbf
		{\bibinfo {volume} {9}},\ \bibinfo {pages} {2204558} (\bibinfo {year}
		{2022})}\BibitemShut {NoStop}%
	\bibitem [{\citenamefont {Hong}\ \emph
		{et~al.}(2021{\natexlab{a}})\citenamefont {Hong}, \citenamefont {Ma},
		\citenamefont {Gao}, \citenamefont {Liu}, \citenamefont {Xiao}, \citenamefont
		{Iqbal},\ and\ \citenamefont {Cui}}]{45}%
	\BibitemOpen
	\bibfield  {author} {\bibinfo {author} {\bibfnamefont {Q.~R.}\ \bibnamefont
			{Hong}}, \bibinfo {author} {\bibfnamefont {Q.}~\bibnamefont {Ma}}, \bibinfo
		{author} {\bibfnamefont {X.~X.}\ \bibnamefont {Gao}}, \bibinfo {author}
		{\bibfnamefont {C.}~\bibnamefont {Liu}}, \bibinfo {author} {\bibfnamefont
			{Q.}~\bibnamefont {Xiao}}, \bibinfo {author} {\bibfnamefont {S.}~\bibnamefont
			{Iqbal}},\ and\ \bibinfo {author} {\bibfnamefont {T.~J.}\ \bibnamefont
			{Cui}},\ }\bibfield  {title} {\bibinfo {title} {Programmable amplitude-coding
			metasurface with multifrequency modulations},\ }\href@noop {} {\bibfield
		{journal} {\bibinfo  {journal} {Advanced Intelligent Systems}\ }\textbf
		{\bibinfo {volume} {3}},\ \bibinfo {pages} {2000260} (\bibinfo {year}
		{2021}{\natexlab{a}})}\BibitemShut {NoStop}%
	\bibitem [{\citenamefont {Jwair}\ \emph {et~al.}(2023)\citenamefont {Jwair},
		\citenamefont {Elwi}, \citenamefont {Alibakhshikenari}, \citenamefont
		{Virdee}, \citenamefont {Almizan}, \citenamefont {Hassain}, \citenamefont
		{Ali}, \citenamefont {Kouhalvandi}, \citenamefont {Livreri}, \citenamefont
		{Tokan} \emph {et~al.}}]{62}%
	\BibitemOpen
	\bibfield  {author} {\bibinfo {author} {\bibfnamefont {M.~H.}\ \bibnamefont
			{Jwair}}, \bibinfo {author} {\bibfnamefont {T.~A.}\ \bibnamefont {Elwi}},
		\bibinfo {author} {\bibfnamefont {M.}~\bibnamefont {Alibakhshikenari}},
		\bibinfo {author} {\bibfnamefont {B.~S.}\ \bibnamefont {Virdee}}, \bibinfo
		{author} {\bibfnamefont {H.}~\bibnamefont {Almizan}}, \bibinfo {author}
		{\bibfnamefont {Z.~A.~A.}\ \bibnamefont {Hassain}}, \bibinfo {author}
		{\bibfnamefont {S.~M.}\ \bibnamefont {Ali}}, \bibinfo {author} {\bibfnamefont
			{L.}~\bibnamefont {Kouhalvandi}}, \bibinfo {author} {\bibfnamefont
			{P.}~\bibnamefont {Livreri}}, \bibinfo {author} {\bibfnamefont {N.~T.}\
			\bibnamefont {Tokan}}, \emph {et~al.},\ }\bibfield  {title} {\bibinfo {title}
		{Intelligent metasurface layer for direct antenna amplitude modulation
			scheme},\ }\href@noop {} {\bibfield  {journal} {\bibinfo  {journal} {IEEE
				access}\ } (\bibinfo {year} {2023})}\BibitemShut {NoStop}%
	\bibitem [{\citenamefont {Luo}\ \emph {et~al.}(2021)\citenamefont {Luo},
		\citenamefont {Ruan},\ and\ \citenamefont {Chen}}]{63}%
	\BibitemOpen
	\bibfield  {author} {\bibinfo {author} {\bibfnamefont {S.~S.}\ \bibnamefont
			{Luo}}, \bibinfo {author} {\bibfnamefont {Y.}~\bibnamefont {Ruan}},\ and\
		\bibinfo {author} {\bibfnamefont {L.}~\bibnamefont {Chen}},\ }\bibfield
	{title} {\bibinfo {title} {Optical-transparent metasurface for flexible
			manipulation and analog information modulation},\ }\href@noop {} {\bibfield
		{journal} {\bibinfo  {journal} {Optics Express}\ }\textbf {\bibinfo {volume}
			{29}},\ \bibinfo {pages} {5867} (\bibinfo {year} {2021})}\BibitemShut
	{NoStop}%
	\bibitem [{\citenamefont {Wu}\ \emph {et~al.}(2023)\citenamefont {Wu},
		\citenamefont {He}, \citenamefont {Wu}, \citenamefont {Bao},\ and\
		\citenamefont {Cui}}]{64}%
	\BibitemOpen
	\bibfield  {author} {\bibinfo {author} {\bibfnamefont {R.~Y.}\ \bibnamefont
			{Wu}}, \bibinfo {author} {\bibfnamefont {S.}~\bibnamefont {He}}, \bibinfo
		{author} {\bibfnamefont {J.~W.}\ \bibnamefont {Wu}}, \bibinfo {author}
		{\bibfnamefont {L.}~\bibnamefont {Bao}},\ and\ \bibinfo {author}
		{\bibfnamefont {T.~J.}\ \bibnamefont {Cui}},\ }\bibfield  {title} {\bibinfo
		{title} {Multi-frequency amplitude-programmable metasurface for multi-channel
			electromagnetic controls},\ }\href@noop {} {\bibfield  {journal} {\bibinfo
			{journal} {Nanophotonics}\ }\textbf {\bibinfo {volume} {12}},\ \bibinfo
		{pages} {2433} (\bibinfo {year} {2023})}\BibitemShut {NoStop}%
	\bibitem [{\citenamefont {Hong}\ \emph
		{et~al.}(2021{\natexlab{b}})\citenamefont {Hong}, \citenamefont {Ma},
		\citenamefont {Gao}, \citenamefont {Liu}, \citenamefont {Xiao}, \citenamefont
		{Iqbal},\ and\ \citenamefont {Cui}}]{65}%
	\BibitemOpen
	\bibfield  {author} {\bibinfo {author} {\bibfnamefont {Q.~R.}\ \bibnamefont
			{Hong}}, \bibinfo {author} {\bibfnamefont {Q.}~\bibnamefont {Ma}}, \bibinfo
		{author} {\bibfnamefont {X.~X.}\ \bibnamefont {Gao}}, \bibinfo {author}
		{\bibfnamefont {C.}~\bibnamefont {Liu}}, \bibinfo {author} {\bibfnamefont
			{Q.}~\bibnamefont {Xiao}}, \bibinfo {author} {\bibfnamefont {S.}~\bibnamefont
			{Iqbal}},\ and\ \bibinfo {author} {\bibfnamefont {T.~J.}\ \bibnamefont
			{Cui}},\ }\bibfield  {title} {\bibinfo {title} {Programmable amplitude-coding
			metasurface with multifrequency modulations},\ }\href@noop {} {\bibfield
		{journal} {\bibinfo  {journal} {Advanced Intelligent Systems}\ }\textbf
		{\bibinfo {volume} {3}},\ \bibinfo {pages} {2000260} (\bibinfo {year}
		{2021}{\natexlab{b}})}\BibitemShut {NoStop}%
	\bibitem [{\citenamefont {Li}\ \emph {et~al.}(2015)\citenamefont {Li},
		\citenamefont {Li}, \citenamefont {Zhang}, \citenamefont {Song},\ and\
		\citenamefont {Liu}}]{46}%
	\BibitemOpen
	\bibfield  {author} {\bibinfo {author} {\bibfnamefont {M.}~\bibnamefont
			{Li}}, \bibinfo {author} {\bibfnamefont {T.}~\bibnamefont {Li}}, \bibinfo
		{author} {\bibfnamefont {X.}~\bibnamefont {Zhang}}, \bibinfo {author}
		{\bibfnamefont {Y.}~\bibnamefont {Song}},\ and\ \bibinfo {author}
		{\bibfnamefont {Y.}~\bibnamefont {Liu}},\ }\bibfield  {title} {\bibinfo
		{title} {Bit error rate analysis of ook modulation scheme under non-coherent
			demodulation for space uplink optical communication systems},\ }in\
	\href@noop {} {\emph {\bibinfo {booktitle} {2015 IEEE 16th international
				conference on communication technology (ICCT)}}}\ (\bibinfo {organization}
	{IEEE},\ \bibinfo {year} {2015})\ pp.\ \bibinfo {pages}
	{695--698}\BibitemShut {NoStop}%
	\bibitem [{\citenamefont {Ke}\ \emph {et~al.}(2021)\citenamefont {Ke},
		\citenamefont {Jiang}, \citenamefont {Chen}, \citenamefont {Zhang},\ and\
		\citenamefont {Cheng}}]{47}%
	\BibitemOpen
	\bibfield  {author} {\bibinfo {author} {\bibfnamefont {J.~C.}\ \bibnamefont
			{Ke}}, \bibinfo {author} {\bibfnamefont {Y.~X.}\ \bibnamefont {Jiang}},
		\bibinfo {author} {\bibfnamefont {M.~Z.}\ \bibnamefont {Chen}}, \bibinfo
		{author} {\bibfnamefont {J.~W.}\ \bibnamefont {Zhang}},\ and\ \bibinfo
		{author} {\bibfnamefont {Q.}~\bibnamefont {Cheng}},\ }\bibfield  {title}
	{\bibinfo {title} {An ook wireless communication system based on transmissive
			digital coding metasurface},\ }in\ \href@noop {} {\emph {\bibinfo {booktitle}
			{2021 Cross Strait Radio Science and Wireless Technology Conference
				(CSRSWTC)}}}\ (\bibinfo {organization} {IEEE},\ \bibinfo {year} {2021})\ pp.\
	\bibinfo {pages} {86--88}\BibitemShut {NoStop}%
	\bibitem [{\citenamefont {Zaiton}\ \emph {et~al.}(2022)\citenamefont {Zaiton},
		\citenamefont {Joyce}, \citenamefont {Ahmad}, \citenamefont {Jasman},\ and\
		\citenamefont {Hassan}}]{48}%
	\BibitemOpen
	\bibfield  {author} {\bibinfo {author} {\bibfnamefont {A.}~\bibnamefont
			{Zaiton}}, \bibinfo {author} {\bibfnamefont {Y.}~\bibnamefont {Joyce}},
		\bibinfo {author} {\bibfnamefont {Z.}~\bibnamefont {Ahmad}}, \bibinfo
		{author} {\bibfnamefont {F.}~\bibnamefont {Jasman}},\ and\ \bibinfo {author}
		{\bibfnamefont {W.}~\bibnamefont {Hassan}},\ }\bibfield  {title} {\bibinfo
		{title} {Performance evaluation of nrz-ook and carrier-less amplitude phase
			modulation in li-fi environment},\ }in\ \href@noop {} {\emph {\bibinfo
			{booktitle} {Journal of Physics: Conference Series}}},\ Vol.\ \bibinfo
	{volume} {2411}\ (\bibinfo {organization} {IOP Publishing},\ \bibinfo {year}
	{2022})\ p.\ \bibinfo {pages} {012016}\BibitemShut {NoStop}%
	\bibitem [{\citenamefont {Khalid}\ \emph {et~al.}(2017)\citenamefont {Khalid},
		\citenamefont {Abbasi},\ and\ \citenamefont {Akan}}]{66}%
	\BibitemOpen
	\bibfield  {author} {\bibinfo {author} {\bibfnamefont {N.}~\bibnamefont
			{Khalid}}, \bibinfo {author} {\bibfnamefont {N.~A.}\ \bibnamefont {Abbasi}},\
		and\ \bibinfo {author} {\bibfnamefont {O.~B.}\ \bibnamefont {Akan}},\
	}\bibfield  {title} {\bibinfo {title} {Capacity and coverage analysis for
			fd-mimo based thz band 5g indoor internet of things},\ }in\ \href@noop {}
	{\emph {\bibinfo {booktitle} {2017 IEEE 28th Annual International Symposium
				on Personal, Indoor, and Mobile Radio Communications (PIMRC)}}}\ (\bibinfo
	{organization} {IEEE},\ \bibinfo {year} {2017})\ pp.\ \bibinfo {pages}
	{1--7}\BibitemShut {NoStop}%
	\bibitem [{\citenamefont {Rouhi}\ \emph {et~al.}(2019)\citenamefont {Rouhi},
		\citenamefont {Rajabalipanah},\ and\ \citenamefont
		{Abdolali}}]{rouhi2019multi}%
	\BibitemOpen
	\bibfield  {author} {\bibinfo {author} {\bibfnamefont {K.}~\bibnamefont
			{Rouhi}}, \bibinfo {author} {\bibfnamefont {H.}~\bibnamefont
			{Rajabalipanah}},\ and\ \bibinfo {author} {\bibfnamefont {A.}~\bibnamefont
			{Abdolali}},\ }\bibfield  {title} {\bibinfo {title} {Multi-bit graphene-based
			bias-encoded metasurfaces for real-time terahertz wavefront shaping: From
			controllable orbital angular momentum generation toward arbitrary beam
			tailoring},\ }\href@noop {} {\bibfield  {journal} {\bibinfo  {journal}
			{Carbon}\ }\textbf {\bibinfo {volume} {149}},\ \bibinfo {pages} {125}
		(\bibinfo {year} {2019})}\BibitemShut {NoStop}%
	\bibitem [{\citenamefont {Tahmasebi}\ \emph {et~al.}(2022)\citenamefont
		{Tahmasebi}, \citenamefont {Abdolali}, \citenamefont {Rajabalipanah},
		\citenamefont {Momeni},\ and\ \citenamefont
		{Fleury}}]{tahmasebi2022parallel}%
	\BibitemOpen
	\bibfield  {author} {\bibinfo {author} {\bibfnamefont {O.}~\bibnamefont
			{Tahmasebi}}, \bibinfo {author} {\bibfnamefont {A.}~\bibnamefont {Abdolali}},
		\bibinfo {author} {\bibfnamefont {H.}~\bibnamefont {Rajabalipanah}}, \bibinfo
		{author} {\bibfnamefont {A.}~\bibnamefont {Momeni}},\ and\ \bibinfo {author}
		{\bibfnamefont {R.}~\bibnamefont {Fleury}},\ }\bibfield  {title} {\bibinfo
		{title} {Parallel temporal signal processing enabled by
			polarization-multiplexed programmable thz metasurfaces},\ }\href@noop {}
	{\bibfield  {journal} {\bibinfo  {journal} {Optics Express}\ }\textbf
		{\bibinfo {volume} {30}},\ \bibinfo {pages} {45221} (\bibinfo {year}
		{2022})}\BibitemShut {NoStop}%
	\bibitem [{\citenamefont {Farzin}\ and\ \citenamefont {Soleimani}(2023)}]{49}%
	\BibitemOpen
	\bibfield  {author} {\bibinfo {author} {\bibfnamefont {P.}~\bibnamefont
			{Farzin}}\ and\ \bibinfo {author} {\bibfnamefont {M.}~\bibnamefont
			{Soleimani}},\ }\bibfield  {title} {\bibinfo {title} {Graphene-based
			metasurface for real-time control of three electromagnetic wave modes and
			polarization state},\ }\href@noop {} {\bibfield  {journal} {\bibinfo
			{journal} {Diamond and Related Materials}\ }\textbf {\bibinfo {volume}
			{139}},\ \bibinfo {pages} {110279} (\bibinfo {year} {2023})}\BibitemShut
	{NoStop}%
	\bibitem [{\citenamefont {John}\ \emph {et~al.}(2020)\citenamefont {John},
		\citenamefont {Gutierrez}, \citenamefont {Zhang}, \citenamefont {Karl},
		\citenamefont {Ramanathan}, \citenamefont {Orobtchouk}, \citenamefont
		{Moreno},\ and\ \citenamefont {Cueff}}]{john2020multipolar}%
	\BibitemOpen
	\bibfield  {author} {\bibinfo {author} {\bibfnamefont {J.}~\bibnamefont
			{John}}, \bibinfo {author} {\bibfnamefont {Y.}~\bibnamefont {Gutierrez}},
		\bibinfo {author} {\bibfnamefont {Z.}~\bibnamefont {Zhang}}, \bibinfo
		{author} {\bibfnamefont {H.}~\bibnamefont {Karl}}, \bibinfo {author}
		{\bibfnamefont {S.}~\bibnamefont {Ramanathan}}, \bibinfo {author}
		{\bibfnamefont {R.}~\bibnamefont {Orobtchouk}}, \bibinfo {author}
		{\bibfnamefont {F.}~\bibnamefont {Moreno}},\ and\ \bibinfo {author}
		{\bibfnamefont {S.}~\bibnamefont {Cueff}},\ }\bibfield  {title} {\bibinfo
		{title} {Multipolar resonances with designer tunability using vo 2
			phase-change materials},\ }\href@noop {} {\bibfield  {journal} {\bibinfo
			{journal} {Physical Review Applied}\ }\textbf {\bibinfo {volume} {13}},\
		\bibinfo {pages} {044053} (\bibinfo {year} {2020})}\BibitemShut {NoStop}%
	\bibitem [{\citenamefont {Kargar}\ \emph {et~al.}(2020)\citenamefont {Kargar},
		\citenamefont {Rouhi},\ and\ \citenamefont
		{Abdolali}}]{kargar2020reprogrammable}%
	\BibitemOpen
	\bibfield  {author} {\bibinfo {author} {\bibfnamefont {R.}~\bibnamefont
			{Kargar}}, \bibinfo {author} {\bibfnamefont {K.}~\bibnamefont {Rouhi}},\ and\
		\bibinfo {author} {\bibfnamefont {A.}~\bibnamefont {Abdolali}},\ }\bibfield
	{title} {\bibinfo {title} {Reprogrammable multifocal thz metalens based on
			metal--insulator transition of vo2-assisted digital metasurface},\
	}\href@noop {} {\bibfield  {journal} {\bibinfo  {journal} {Optics
				Communications}\ }\textbf {\bibinfo {volume} {462}},\ \bibinfo {pages}
		{125331} (\bibinfo {year} {2020})}\BibitemShut {NoStop}%
	\bibitem [{\citenamefont {Isi{\'c}}\ \emph {et~al.}(2015)\citenamefont
		{Isi{\'c}}, \citenamefont {Vasi{\'c}}, \citenamefont {Zografopoulos},
		\citenamefont {Beccherelli},\ and\ \citenamefont {Gaji{\'c}}}]{52}%
	\BibitemOpen
	\bibfield  {author} {\bibinfo {author} {\bibfnamefont {G.}~\bibnamefont
			{Isi{\'c}}}, \bibinfo {author} {\bibfnamefont {B.}~\bibnamefont {Vasi{\'c}}},
		\bibinfo {author} {\bibfnamefont {D.~C.}\ \bibnamefont {Zografopoulos}},
		\bibinfo {author} {\bibfnamefont {R.}~\bibnamefont {Beccherelli}},\ and\
		\bibinfo {author} {\bibfnamefont {R.}~\bibnamefont {Gaji{\'c}}},\ }\bibfield
	{title} {\bibinfo {title} {Electrically tunable critically coupled terahertz
			metamaterial absorber based on nematic liquid crystals},\ }\href@noop {}
	{\bibfield  {journal} {\bibinfo  {journal} {Physical Review Applied}\
		}\textbf {\bibinfo {volume} {3}},\ \bibinfo {pages} {064007} (\bibinfo {year}
		{2015})}\BibitemShut {NoStop}%
	\bibitem [{\citenamefont {Soleimani}\ \emph {et~al.}(2022)\citenamefont
		{Soleimani}, \citenamefont {Rouhi},\ and\ \citenamefont
		{Momeni}}]{soleimani2022near}%
	\BibitemOpen
	\bibfield  {author} {\bibinfo {author} {\bibfnamefont {S.}~\bibnamefont
			{Soleimani}}, \bibinfo {author} {\bibfnamefont {K.}~\bibnamefont {Rouhi}},\
		and\ \bibinfo {author} {\bibfnamefont {A.}~\bibnamefont {Momeni}},\
	}\bibfield  {title} {\bibinfo {title} {Near-field optical mimo communication
			with polarization-dependent metasurfaces},\ }\href@noop {} {\bibfield
		{journal} {\bibinfo  {journal} {arXiv preprint arXiv:2210.10237}\ } (\bibinfo
		{year} {2022})}\BibitemShut {NoStop}%
	\bibitem [{\citenamefont {Li}\ \emph {et~al.}(2022)\citenamefont {Li},
		\citenamefont {Krisshnamurthi}, \citenamefont {Luo}, \citenamefont {Swan},
		\citenamefont {Ling},\ and\ \citenamefont {Paiella}}]{55}%
	\BibitemOpen
	\bibfield  {author} {\bibinfo {author} {\bibfnamefont {Y.}~\bibnamefont
			{Li}}, \bibinfo {author} {\bibfnamefont {M.~R.}\ \bibnamefont
			{Krisshnamurthi}}, \bibinfo {author} {\bibfnamefont {W.}~\bibnamefont {Luo}},
		\bibinfo {author} {\bibfnamefont {A.~K.}\ \bibnamefont {Swan}}, \bibinfo
		{author} {\bibfnamefont {X.}~\bibnamefont {Ling}},\ and\ \bibinfo {author}
		{\bibfnamefont {R.}~\bibnamefont {Paiella}},\ }\bibfield  {title} {\bibinfo
		{title} {Graphene metasurfaces for terahertz wavefront shaping and light
			emission},\ }\href@noop {} {\bibfield  {journal} {\bibinfo  {journal}
			{Optical Materials Express}\ }\textbf {\bibinfo {volume} {12}},\ \bibinfo
		{pages} {4528} (\bibinfo {year} {2022})}\BibitemShut {NoStop}%
	\bibitem [{\citenamefont {Xiao}\ \emph {et~al.}(2020)\citenamefont {Xiao},
		\citenamefont {Zhang}, \citenamefont {Tong}, \citenamefont {Yu},\ and\
		\citenamefont {Xiao}}]{56}%
	\BibitemOpen
	\bibfield  {author} {\bibinfo {author} {\bibfnamefont {B.}~\bibnamefont
			{Xiao}}, \bibinfo {author} {\bibfnamefont {Y.}~\bibnamefont {Zhang}},
		\bibinfo {author} {\bibfnamefont {S.}~\bibnamefont {Tong}}, \bibinfo {author}
		{\bibfnamefont {J.}~\bibnamefont {Yu}},\ and\ \bibinfo {author}
		{\bibfnamefont {L.}~\bibnamefont {Xiao}},\ }\bibfield  {title} {\bibinfo
		{title} {Novel tunable graphene-encoded metasurfaces on an uneven substrate
			for beam-steering in far-field at the terahertz frequencies},\ }\href@noop {}
	{\bibfield  {journal} {\bibinfo  {journal} {Optics Express}\ }\textbf
		{\bibinfo {volume} {28}},\ \bibinfo {pages} {7125} (\bibinfo {year}
		{2020})}\BibitemShut {NoStop}%
	\bibitem [{\citenamefont {Lan}\ \emph {et~al.}(2023)\citenamefont {Lan},
		\citenamefont {Wang}, \citenamefont {Zeng}, \citenamefont {Liang},
		\citenamefont {Song}, \citenamefont {Liu}, \citenamefont {Mazumder},
		\citenamefont {Yang}, \citenamefont {Zhang},\ and\ \citenamefont
		{Mittleman}}]{57}%
	\BibitemOpen
	\bibfield  {author} {\bibinfo {author} {\bibfnamefont {F.}~\bibnamefont
			{Lan}}, \bibinfo {author} {\bibfnamefont {L.}~\bibnamefont {Wang}}, \bibinfo
		{author} {\bibfnamefont {H.}~\bibnamefont {Zeng}}, \bibinfo {author}
		{\bibfnamefont {S.}~\bibnamefont {Liang}}, \bibinfo {author} {\bibfnamefont
			{T.}~\bibnamefont {Song}}, \bibinfo {author} {\bibfnamefont {W.}~\bibnamefont
			{Liu}}, \bibinfo {author} {\bibfnamefont {P.}~\bibnamefont {Mazumder}},
		\bibinfo {author} {\bibfnamefont {Z.}~\bibnamefont {Yang}}, \bibinfo {author}
		{\bibfnamefont {Y.}~\bibnamefont {Zhang}},\ and\ \bibinfo {author}
		{\bibfnamefont {D.~M.}\ \bibnamefont {Mittleman}},\ }\bibfield  {title}
	{\bibinfo {title} {Real-time programmable metasurface for terahertz
			multifunctional wave front engineering},\ }\href@noop {} {\bibfield
		{journal} {\bibinfo  {journal} {Light: Science \& Applications}\ }\textbf
		{\bibinfo {volume} {12}},\ \bibinfo {pages} {191} (\bibinfo {year}
		{2023})}\BibitemShut {NoStop}%
	\bibitem [{\citenamefont {Yang}\ \emph {et~al.}(2024)\citenamefont {Yang},
		\citenamefont {Tan}, \citenamefont {Prakash}, \citenamefont {Kumar},
		\citenamefont {Ariando}, \citenamefont {Singh}, \citenamefont {Wang},\ and\
		\citenamefont {Pitchappa}}]{58}%
	\BibitemOpen
	\bibfield  {author} {\bibinfo {author} {\bibfnamefont {F.}~\bibnamefont
			{Yang}}, \bibinfo {author} {\bibfnamefont {T.~C.}\ \bibnamefont {Tan}},
		\bibinfo {author} {\bibfnamefont {S.}~\bibnamefont {Prakash}}, \bibinfo
		{author} {\bibfnamefont {A.}~\bibnamefont {Kumar}}, \bibinfo {author}
		{\bibfnamefont {A.}~\bibnamefont {Ariando}}, \bibinfo {author} {\bibfnamefont
			{R.}~\bibnamefont {Singh}}, \bibinfo {author} {\bibfnamefont
			{N.}~\bibnamefont {Wang}},\ and\ \bibinfo {author} {\bibfnamefont
			{P.}~\bibnamefont {Pitchappa}},\ }\bibfield  {title} {\bibinfo {title}
		{Reconfigurable wide-angle beam-steering terahertz metasurfaces based on
			vanadium dioxide},\ }\href@noop {} {\bibfield  {journal} {\bibinfo  {journal}
			{Advanced Optical Materials}\ }\textbf {\bibinfo {volume} {12}},\ \bibinfo
		{pages} {2302047} (\bibinfo {year} {2024})}\BibitemShut {NoStop}%
	\bibitem [{\citenamefont {Li}\ and\ \citenamefont {Paiella}(2021)}]{59}%
	\BibitemOpen
	\bibfield  {author} {\bibinfo {author} {\bibfnamefont {Y.}~\bibnamefont
			{Li}}\ and\ \bibinfo {author} {\bibfnamefont {R.}~\bibnamefont {Paiella}},\
	}\bibfield  {title} {\bibinfo {title} {Tunable terahertz metasurface platform
			based on cvd graphene plasmonics},\ }\href@noop {} {\bibfield  {journal}
		{\bibinfo  {journal} {Optics Express}\ }\textbf {\bibinfo {volume} {29}},\
		\bibinfo {pages} {40594} (\bibinfo {year} {2021})}\BibitemShut {NoStop}%
	\bibitem [{\citenamefont {Abed}\ and\ \citenamefont {Yousefi}(2020)}]{60}%
	\BibitemOpen
	\bibfield  {author} {\bibinfo {author} {\bibfnamefont {O.}~\bibnamefont
			{Abed}}\ and\ \bibinfo {author} {\bibfnamefont {L.}~\bibnamefont {Yousefi}},\
	}\bibfield  {title} {\bibinfo {title} {Tunable metasurfaces using phase
			change materials and transparent graphene heaters},\ }\href@noop {}
	{\bibfield  {journal} {\bibinfo  {journal} {Optics express}\ }\textbf
		{\bibinfo {volume} {28}},\ \bibinfo {pages} {33876} (\bibinfo {year}
		{2020})}\BibitemShut {NoStop}%
	\bibitem [{\citenamefont {Fu}\ \emph {et~al.}(2022)\citenamefont {Fu},
		\citenamefont {Shi}, \citenamefont {Yang}, \citenamefont {Fu}, \citenamefont
		{Liu}, \citenamefont {Wu}, \citenamefont {Yang}, \citenamefont {Bao},\ and\
		\citenamefont {Cui}}]{61}%
	\BibitemOpen
	\bibfield  {author} {\bibinfo {author} {\bibfnamefont {X.}~\bibnamefont
			{Fu}}, \bibinfo {author} {\bibfnamefont {L.}~\bibnamefont {Shi}}, \bibinfo
		{author} {\bibfnamefont {J.}~\bibnamefont {Yang}}, \bibinfo {author}
		{\bibfnamefont {Y.}~\bibnamefont {Fu}}, \bibinfo {author} {\bibfnamefont
			{C.}~\bibnamefont {Liu}}, \bibinfo {author} {\bibfnamefont {J.~W.}\
			\bibnamefont {Wu}}, \bibinfo {author} {\bibfnamefont {F.}~\bibnamefont
			{Yang}}, \bibinfo {author} {\bibfnamefont {L.}~\bibnamefont {Bao}},\ and\
		\bibinfo {author} {\bibfnamefont {T.~J.}\ \bibnamefont {Cui}},\ }\bibfield
	{title} {\bibinfo {title} {Flexible terahertz beam manipulations based on
			liquid-crystal-integrated programmable metasurfaces},\ }\href@noop {}
	{\bibfield  {journal} {\bibinfo  {journal} {ACS Applied Materials \&
				Interfaces}\ }\textbf {\bibinfo {volume} {14}},\ \bibinfo {pages} {22287}
		(\bibinfo {year} {2022})}\BibitemShut {NoStop}%
	\bibitem [{\citenamefont {Low}\ and\ \citenamefont
		{Avouris}(2014)}]{low2014graphene}%
	\BibitemOpen
	\bibfield  {author} {\bibinfo {author} {\bibfnamefont {T.}~\bibnamefont
			{Low}}\ and\ \bibinfo {author} {\bibfnamefont {P.}~\bibnamefont {Avouris}},\
	}\bibfield  {title} {\bibinfo {title} {Graphene plasmonics for terahertz to
			mid-infrared applications},\ }\href@noop {} {\bibfield  {journal} {\bibinfo
			{journal} {ACS nano}\ }\textbf {\bibinfo {volume} {8}},\ \bibinfo {pages}
		{1086} (\bibinfo {year} {2014})}\BibitemShut {NoStop}%
	\bibitem [{\citenamefont {Hanson}(2008)}]{hanson2008dyadic}%
	\BibitemOpen
	\bibfield  {author} {\bibinfo {author} {\bibfnamefont {G.~W.}\ \bibnamefont
			{Hanson}},\ }\bibfield  {title} {\bibinfo {title} {Dyadic green’s functions
			and guided surface waves for a surface conductivity model of graphene},\
	}\href@noop {} {\bibfield  {journal} {\bibinfo  {journal} {Journal of Applied
				Physics}\ }\textbf {\bibinfo {volume} {103}} (\bibinfo {year}
		{2008})}\BibitemShut {NoStop}%
	\bibitem [{\citenamefont {Vasi{\'c}}\ \emph {et~al.}(2013)\citenamefont
		{Vasi{\'c}}, \citenamefont {Jakovljevi{\'c}}, \citenamefont {Isi{\'c}},\ and\
		\citenamefont {Gaji{\'c}}}]{vasic2013tunable}%
	\BibitemOpen
	\bibfield  {author} {\bibinfo {author} {\bibfnamefont {B.}~\bibnamefont
			{Vasi{\'c}}}, \bibinfo {author} {\bibfnamefont {M.~M.}\ \bibnamefont
			{Jakovljevi{\'c}}}, \bibinfo {author} {\bibfnamefont {G.}~\bibnamefont
			{Isi{\'c}}},\ and\ \bibinfo {author} {\bibfnamefont {R.}~\bibnamefont
			{Gaji{\'c}}},\ }\bibfield  {title} {\bibinfo {title} {Tunable metamaterials
			based on split ring resonators and doped graphene},\ }\href@noop {}
	{\bibfield  {journal} {\bibinfo  {journal} {Applied Physics Letters}\
		}\textbf {\bibinfo {volume} {103}} (\bibinfo {year} {2013})}\BibitemShut
	{NoStop}%
	\bibitem [{\citenamefont {Novoselov}\ \emph {et~al.}(2004)\citenamefont
		{Novoselov}, \citenamefont {Geim}, \citenamefont {Morozov}, \citenamefont
		{Jiang}, \citenamefont {Zhang}, \citenamefont {Dubonos}, \citenamefont
		{Grigorieva},\ and\ \citenamefont {Firsov}}]{novoselov2004electric}%
	\BibitemOpen
	\bibfield  {author} {\bibinfo {author} {\bibfnamefont {K.~S.}\ \bibnamefont
			{Novoselov}}, \bibinfo {author} {\bibfnamefont {A.~K.}\ \bibnamefont {Geim}},
		\bibinfo {author} {\bibfnamefont {S.~V.}\ \bibnamefont {Morozov}}, \bibinfo
		{author} {\bibfnamefont {D.-e.}\ \bibnamefont {Jiang}}, \bibinfo {author}
		{\bibfnamefont {Y.}~\bibnamefont {Zhang}}, \bibinfo {author} {\bibfnamefont
			{S.~V.}\ \bibnamefont {Dubonos}}, \bibinfo {author} {\bibfnamefont {I.~V.}\
			\bibnamefont {Grigorieva}},\ and\ \bibinfo {author} {\bibfnamefont {A.~A.}\
			\bibnamefont {Firsov}},\ }\bibfield  {title} {\bibinfo {title} {Electric
			field effect in atomically thin carbon films},\ }\href@noop {} {\bibfield
		{journal} {\bibinfo  {journal} {science}\ }\textbf {\bibinfo {volume}
			{306}},\ \bibinfo {pages} {666} (\bibinfo {year} {2004})}\BibitemShut
	{NoStop}%
	\bibitem [{\citenamefont {Ju}\ \emph {et~al.}(2011)\citenamefont {Ju},
		\citenamefont {Geng}, \citenamefont {Horng}, \citenamefont {Girit},
		\citenamefont {Martin}, \citenamefont {Hao}, \citenamefont {Bechtel},
		\citenamefont {Liang}, \citenamefont {Zettl}, \citenamefont {Shen} \emph
		{et~al.}}]{ju2011graphene}%
	\BibitemOpen
	\bibfield  {author} {\bibinfo {author} {\bibfnamefont {L.}~\bibnamefont
			{Ju}}, \bibinfo {author} {\bibfnamefont {B.}~\bibnamefont {Geng}}, \bibinfo
		{author} {\bibfnamefont {J.}~\bibnamefont {Horng}}, \bibinfo {author}
		{\bibfnamefont {C.}~\bibnamefont {Girit}}, \bibinfo {author} {\bibfnamefont
			{M.}~\bibnamefont {Martin}}, \bibinfo {author} {\bibfnamefont
			{Z.}~\bibnamefont {Hao}}, \bibinfo {author} {\bibfnamefont {H.~A.}\
			\bibnamefont {Bechtel}}, \bibinfo {author} {\bibfnamefont {X.}~\bibnamefont
			{Liang}}, \bibinfo {author} {\bibfnamefont {A.}~\bibnamefont {Zettl}},
		\bibinfo {author} {\bibfnamefont {Y.~R.}\ \bibnamefont {Shen}}, \emph
		{et~al.},\ }\bibfield  {title} {\bibinfo {title} {Graphene plasmonics for
			tunable terahertz metamaterials},\ }\href@noop {} {\bibfield  {journal}
		{\bibinfo  {journal} {Nature nanotechnology}\ }\textbf {\bibinfo {volume}
			{6}},\ \bibinfo {pages} {630} (\bibinfo {year} {2011})}\BibitemShut {NoStop}%
	\bibitem [{\citenamefont {Rouhi}\ \emph {et~al.}(2021)\citenamefont {Rouhi},
		\citenamefont {Abdolali},\ and\ \citenamefont {Fallah}}]{rouhi2021designing}%
	\BibitemOpen
	\bibfield  {author} {\bibinfo {author} {\bibfnamefont {K.}~\bibnamefont
			{Rouhi}}, \bibinfo {author} {\bibfnamefont {A.}~\bibnamefont {Abdolali}},\
		and\ \bibinfo {author} {\bibfnamefont {S.}~\bibnamefont {Fallah}},\
	}\bibfield  {title} {\bibinfo {title} {Designing approach of terahertz
			broadband backscattering reduction based on combination of diffusion and
			absorption},\ }\href@noop {} {\bibfield  {journal} {\bibinfo  {journal}
			{Optik}\ }\textbf {\bibinfo {volume} {246}},\ \bibinfo {pages} {167771}
		(\bibinfo {year} {2021})}\BibitemShut {NoStop}%
	\bibitem [{\citenamefont {Liu}\ \emph {et~al.}(2019)\citenamefont {Liu},
		\citenamefont {Guo},\ and\ \citenamefont {Zhang}}]{liu2019simple}%
	\BibitemOpen
	\bibfield  {author} {\bibinfo {author} {\bibfnamefont {Z.}~\bibnamefont
			{Liu}}, \bibinfo {author} {\bibfnamefont {L.}~\bibnamefont {Guo}},\ and\
		\bibinfo {author} {\bibfnamefont {Q.}~\bibnamefont {Zhang}},\ }\bibfield
	{title} {\bibinfo {title} {A simple and efficient method for designing
			broadband terahertz absorber based on singular graphene metasurface},\
	}\href@noop {} {\bibfield  {journal} {\bibinfo  {journal} {Nanomaterials}\
		}\textbf {\bibinfo {volume} {9}},\ \bibinfo {pages} {1351} (\bibinfo {year}
		{2019})}\BibitemShut {NoStop}%
	\bibitem [{\citenamefont {Pae}\ \emph {et~al.}(2022)\citenamefont {Pae},
		\citenamefont {Im}, \citenamefont {Han},\ and\ \citenamefont {Song}}]{11}%
	\BibitemOpen
	\bibfield  {author} {\bibinfo {author} {\bibfnamefont {J.-S.}\ \bibnamefont
			{Pae}}, \bibinfo {author} {\bibfnamefont {S.-J.}\ \bibnamefont {Im}},
		\bibinfo {author} {\bibfnamefont {Y.-H.}\ \bibnamefont {Han}},\ and\ \bibinfo
		{author} {\bibfnamefont {K.-S.}\ \bibnamefont {Song}},\ }\bibfield  {title}
	{\bibinfo {title} {Analysis and design of a single layer double square slot
			frequency selective surface with single stopband between double passbands},\
	}\href@noop {} {\bibfield  {journal} {\bibinfo  {journal} {Engineering
				Reports}\ }\textbf {\bibinfo {volume} {4}},\ \bibinfo {pages} {e12543}
		(\bibinfo {year} {2022})}\BibitemShut {NoStop}%
	\bibitem [{\citenamefont {Wang}\ \emph {et~al.}(2019)\citenamefont {Wang},
		\citenamefont {Ren}, \citenamefont {Yan}, \citenamefont {Jiang},
		\citenamefont {Sha},\ and\ \citenamefont {Shan}}]{12}%
	\BibitemOpen
	\bibfield  {author} {\bibinfo {author} {\bibfnamefont {R.}~\bibnamefont
			{Wang}}, \bibinfo {author} {\bibfnamefont {X.-G.}\ \bibnamefont {Ren}},
		\bibinfo {author} {\bibfnamefont {Z.}~\bibnamefont {Yan}}, \bibinfo {author}
		{\bibfnamefont {L.-J.}\ \bibnamefont {Jiang}}, \bibinfo {author}
		{\bibfnamefont {W.~E.}\ \bibnamefont {Sha}},\ and\ \bibinfo {author}
		{\bibfnamefont {G.-C.}\ \bibnamefont {Shan}},\ }\bibfield  {title} {\bibinfo
		{title} {Graphene based functional devices: A short review},\ }\href@noop {}
	{\bibfield  {journal} {\bibinfo  {journal} {Frontiers of Physics}\ }\textbf
		{\bibinfo {volume} {14}},\ \bibinfo {pages} {1} (\bibinfo {year}
		{2019})}\BibitemShut {NoStop}%
	\bibitem [{\citenamefont {Dean}\ \emph {et~al.}(2010)\citenamefont {Dean},
		\citenamefont {Young}, \citenamefont {Meric}, \citenamefont {Lee},
		\citenamefont {Wang}, \citenamefont {Sorgenfrei}, \citenamefont {Watanabe},
		\citenamefont {Taniguchi}, \citenamefont {Kim}, \citenamefont {Shepard} \emph
		{et~al.}}]{13}%
	\BibitemOpen
	\bibfield  {author} {\bibinfo {author} {\bibfnamefont {C.~R.}\ \bibnamefont
			{Dean}}, \bibinfo {author} {\bibfnamefont {A.~F.}\ \bibnamefont {Young}},
		\bibinfo {author} {\bibfnamefont {I.}~\bibnamefont {Meric}}, \bibinfo
		{author} {\bibfnamefont {C.}~\bibnamefont {Lee}}, \bibinfo {author}
		{\bibfnamefont {L.}~\bibnamefont {Wang}}, \bibinfo {author} {\bibfnamefont
			{S.}~\bibnamefont {Sorgenfrei}}, \bibinfo {author} {\bibfnamefont
			{K.}~\bibnamefont {Watanabe}}, \bibinfo {author} {\bibfnamefont
			{T.}~\bibnamefont {Taniguchi}}, \bibinfo {author} {\bibfnamefont
			{P.}~\bibnamefont {Kim}}, \bibinfo {author} {\bibfnamefont {K.~L.}\
			\bibnamefont {Shepard}}, \emph {et~al.},\ }\bibfield  {title} {\bibinfo
		{title} {Boron nitride substrates for high-quality graphene electronics},\
	}\href@noop {} {\bibfield  {journal} {\bibinfo  {journal} {Nature
				nanotechnology}\ }\textbf {\bibinfo {volume} {5}},\ \bibinfo {pages} {722}
		(\bibinfo {year} {2010})}\BibitemShut {NoStop}%
	\bibitem [{\citenamefont {Wang}\ and\ \citenamefont
		{Strohbehn}(1974)}]{wang1974perturbed}%
	\BibitemOpen
	\bibfield  {author} {\bibinfo {author} {\bibfnamefont {T.-i.}\ \bibnamefont
			{Wang}}\ and\ \bibinfo {author} {\bibfnamefont {J.~W.}\ \bibnamefont
			{Strohbehn}},\ }\bibfield  {title} {\bibinfo {title} {Perturbed log-normal
			distribution of irradiance fluctuations},\ }\href@noop {} {\bibfield
		{journal} {\bibinfo  {journal} {JOSA}\ }\textbf {\bibinfo {volume} {64}},\
		\bibinfo {pages} {994} (\bibinfo {year} {1974})}\BibitemShut {NoStop}%
	\bibitem [{\citenamefont {Andrews}\ and\ \citenamefont
		{Phillips}(2005)}]{andrews2005laser}%
	\BibitemOpen
	\bibfield  {author} {\bibinfo {author} {\bibfnamefont {L.~C.}\ \bibnamefont
			{Andrews}}\ and\ \bibinfo {author} {\bibfnamefont {R.~L.}\ \bibnamefont
			{Phillips}},\ }\bibfield  {title} {\bibinfo {title} {Laser beam propagation
			through random media},\ }\href@noop {} {\bibfield  {journal} {\bibinfo
			{journal} {Laser Beam Propagation Through Random Media: Second Edition}\ }
		(\bibinfo {year} {2005})}\BibitemShut {NoStop}%
	\bibitem [{\citenamefont {Letzepis}\ and\ \citenamefont
		{Fabregas}(2009)}]{letzepis2009outage}%
	\BibitemOpen
	\bibfield  {author} {\bibinfo {author} {\bibfnamefont {N.}~\bibnamefont
			{Letzepis}}\ and\ \bibinfo {author} {\bibfnamefont {A.~G.~I.}\ \bibnamefont
			{Fabregas}},\ }\bibfield  {title} {\bibinfo {title} {Outage probability of
			the gaussian mimo free-space optical channel with ppm},\ }\href@noop {}
	{\bibfield  {journal} {\bibinfo  {journal} {IEEE Transactions on
				Communications}\ }\textbf {\bibinfo {volume} {57}},\ \bibinfo {pages} {3682}
		(\bibinfo {year} {2009})}\BibitemShut {NoStop}%
	\bibitem [{\citenamefont {Nistazakis}\ \emph {et~al.}(2009)\citenamefont
		{Nistazakis}, \citenamefont {Tsiftsis},\ and\ \citenamefont
		{Tombras}}]{nistazakis2009performance}%
	\BibitemOpen
	\bibfield  {author} {\bibinfo {author} {\bibfnamefont {H.~E.}\ \bibnamefont
			{Nistazakis}}, \bibinfo {author} {\bibfnamefont {T.~A.}\ \bibnamefont
			{Tsiftsis}},\ and\ \bibinfo {author} {\bibfnamefont {G.~S.}\ \bibnamefont
			{Tombras}},\ }\bibfield  {title} {\bibinfo {title} {Performance analysis of
			free-space optical communication systems over atmospheric turbulence
			channels},\ }\href@noop {} {\bibfield  {journal} {\bibinfo  {journal} {IET
				communications}\ }\textbf {\bibinfo {volume} {3}},\ \bibinfo {pages} {1402}
		(\bibinfo {year} {2009})}\BibitemShut {NoStop}%
	\bibitem [{\citenamefont {Osche}(2002)}]{osche2002optical}%
	\BibitemOpen
	\bibfield  {author} {\bibinfo {author} {\bibfnamefont {G.~R.}\ \bibnamefont
			{Osche}},\ }\href@noop {} {\emph {\bibinfo {title} {Optical detection theory
				for laser applications}}}\ (\bibinfo {year} {2002})\BibitemShut {NoStop}%
	\bibitem [{\citenamefont {Popoola}\ \emph {et~al.}(2008)\citenamefont
		{Popoola}, \citenamefont {Ghassemlooy}, \citenamefont {Allen}, \citenamefont
		{Leitgeb},\ and\ \citenamefont {Gao}}]{popoola2008free}%
	\BibitemOpen
	\bibfield  {author} {\bibinfo {author} {\bibfnamefont {W.~O.}\ \bibnamefont
			{Popoola}}, \bibinfo {author} {\bibfnamefont {Z.}~\bibnamefont
			{Ghassemlooy}}, \bibinfo {author} {\bibfnamefont {J.}~\bibnamefont {Allen}},
		\bibinfo {author} {\bibfnamefont {E.}~\bibnamefont {Leitgeb}},\ and\ \bibinfo
		{author} {\bibfnamefont {S.}~\bibnamefont {Gao}},\ }\bibfield  {title}
	{\bibinfo {title} {Free-space optical communication employing subcarrier
			modulation and spatial diversity in atmospheric turbulence channel},\
	}\href@noop {} {\bibfield  {journal} {\bibinfo  {journal} {IET
				optoelectronics}\ }\textbf {\bibinfo {volume} {2}},\ \bibinfo {pages} {16}
		(\bibinfo {year} {2008})}\BibitemShut {NoStop}%
	\bibitem [{\citenamefont {Kwok}\ \emph {et~al.}(2008)\citenamefont {Kwok},
		\citenamefont {Penty},\ and\ \citenamefont {White}}]{kwok2008link}%
	\BibitemOpen
	\bibfield  {author} {\bibinfo {author} {\bibfnamefont {C.}~\bibnamefont
			{Kwok}}, \bibinfo {author} {\bibfnamefont {R.~V.}\ \bibnamefont {Penty}},\
		and\ \bibinfo {author} {\bibfnamefont {I.~H.}\ \bibnamefont {White}},\
	}\bibfield  {title} {\bibinfo {title} {Link reliability improvement for
			optical wireless communication systems with temporal-domain diversity
			reception},\ }\href@noop {} {\bibfield  {journal} {\bibinfo  {journal} {IEEE
				Photonics Technology Letters}\ }\textbf {\bibinfo {volume} {20}},\ \bibinfo
		{pages} {700} (\bibinfo {year} {2008})}\BibitemShut {NoStop}%
	\bibitem [{\citenamefont {Wang}\ \emph {et~al.}(2012)\citenamefont {Wang},
		\citenamefont {Zhong}, \citenamefont {Yu},\ and\ \citenamefont
		{Fu}}]{wang2012performance}%
	\BibitemOpen
	\bibfield  {author} {\bibinfo {author} {\bibfnamefont {Z.}~\bibnamefont
			{Wang}}, \bibinfo {author} {\bibfnamefont {W.-D.}\ \bibnamefont {Zhong}},
		\bibinfo {author} {\bibfnamefont {C.}~\bibnamefont {Yu}},\ and\ \bibinfo
		{author} {\bibfnamefont {S.}~\bibnamefont {Fu}},\ }\bibfield  {title}
	{\bibinfo {title} {Performance improvement of on-off-keying free-space
			optical transmission systems by a co-propagating reference continuous wave
			light},\ }\href@noop {} {\bibfield  {journal} {\bibinfo  {journal} {Optics
				express}\ }\textbf {\bibinfo {volume} {20}},\ \bibinfo {pages} {9284}
		(\bibinfo {year} {2012})}\BibitemShut {NoStop}%
	\bibitem [{\citenamefont {Al-Habash}\ \emph {et~al.}(2001)\citenamefont
		{Al-Habash}, \citenamefont {Andrews},\ and\ \citenamefont
		{Phillips}}]{al2001mathematical}%
	\BibitemOpen
	\bibfield  {author} {\bibinfo {author} {\bibfnamefont {M.}~\bibnamefont
			{Al-Habash}}, \bibinfo {author} {\bibfnamefont {L.~C.}\ \bibnamefont
			{Andrews}},\ and\ \bibinfo {author} {\bibfnamefont {R.~L.}\ \bibnamefont
			{Phillips}},\ }\bibfield  {title} {\bibinfo {title} {Mathematical model for
			the irradiance probability density function of a laser beam propagating
			through turbulent media},\ }\href@noop {} {\bibfield  {journal} {\bibinfo
			{journal} {Optical engineering}\ }\textbf {\bibinfo {volume} {40}},\ \bibinfo
		{pages} {1554} (\bibinfo {year} {2001})}\BibitemShut {NoStop}%
	\bibitem [{\citenamefont {Li}\ \emph {et~al.}(2017)\citenamefont {Li},
		\citenamefont {Ding},\ and\ \citenamefont {Dang}}]{li2017suboptimal}%
	\BibitemOpen
	\bibfield  {author} {\bibinfo {author} {\bibfnamefont {R.}~\bibnamefont
			{Li}}, \bibinfo {author} {\bibfnamefont {S.}~\bibnamefont {Ding}},\ and\
		\bibinfo {author} {\bibfnamefont {A.}~\bibnamefont {Dang}},\ }\bibfield
	{title} {\bibinfo {title} {Suboptimal maximum likelihood detection of on--off
			keying for a wireless optical communication system},\ }\href@noop {}
	{\bibfield  {journal} {\bibinfo  {journal} {JOSA A}\ }\textbf {\bibinfo
			{volume} {34}},\ \bibinfo {pages} {798} (\bibinfo {year} {2017})}\BibitemShut
	{NoStop}%
	\bibitem [{\citenamefont {Zhu}\ and\ \citenamefont {Kahn}(2002)}]{zhu2002free}%
	\BibitemOpen
	\bibfield  {author} {\bibinfo {author} {\bibfnamefont {X.}~\bibnamefont
			{Zhu}}\ and\ \bibinfo {author} {\bibfnamefont {J.~M.}\ \bibnamefont {Kahn}},\
	}\bibfield  {title} {\bibinfo {title} {Free-space optical communication
			through atmospheric turbulence channels},\ }\href@noop {} {\bibfield
		{journal} {\bibinfo  {journal} {IEEE Transactions on communications}\
		}\textbf {\bibinfo {volume} {50}},\ \bibinfo {pages} {1293} (\bibinfo {year}
		{2002})}\BibitemShut {NoStop}%
	\bibitem [{\citenamefont {Proakis}\ and\ \citenamefont
		{Salehi}(2008)}]{proakis2008digital}%
	\BibitemOpen
	\bibfield  {author} {\bibinfo {author} {\bibfnamefont {J.~G.}\ \bibnamefont
			{Proakis}}\ and\ \bibinfo {author} {\bibfnamefont {M.}~\bibnamefont
			{Salehi}},\ }\href@noop {} {\emph {\bibinfo {title} {Digital
				communications}}}\ (\bibinfo  {publisher} {McGraw-hill},\ \bibinfo {year}
	{2008})\BibitemShut {NoStop}%
	\bibitem [{\citenamefont {Borah}\ and\ \citenamefont
		{Voelz}(2009)}]{borah2009pointing}%
	\BibitemOpen
	\bibfield  {author} {\bibinfo {author} {\bibfnamefont {D.~K.}\ \bibnamefont
			{Borah}}\ and\ \bibinfo {author} {\bibfnamefont {D.~G.}\ \bibnamefont
			{Voelz}},\ }\bibfield  {title} {\bibinfo {title} {Pointing error effects on
			free-space optical communication links in the presence of atmospheric
			turbulence},\ }\href@noop {} {\bibfield  {journal} {\bibinfo  {journal}
			{Journal of Lightwave Technology}\ }\textbf {\bibinfo {volume} {27}},\
		\bibinfo {pages} {3965} (\bibinfo {year} {2009})}\BibitemShut {NoStop}%
	\bibitem [{\citenamefont {Li}\ \emph {et~al.}(2007)\citenamefont {Li},
		\citenamefont {Liu},\ and\ \citenamefont {Taylor}}]{li2007optical}%
	\BibitemOpen
	\bibfield  {author} {\bibinfo {author} {\bibfnamefont {J.}~\bibnamefont
			{Li}}, \bibinfo {author} {\bibfnamefont {J.~Q.}\ \bibnamefont {Liu}},\ and\
		\bibinfo {author} {\bibfnamefont {D.~P.}\ \bibnamefont {Taylor}},\ }\bibfield
	{title} {\bibinfo {title} {Optical communication using subcarrier psk
			intensity modulation through atmospheric turbulence channels},\ }\href@noop
	{} {\bibfield  {journal} {\bibinfo  {journal} {IEEE Transactions on
				Communications}\ }\textbf {\bibinfo {volume} {55}},\ \bibinfo {pages} {1598}
		(\bibinfo {year} {2007})}\BibitemShut {NoStop}%
	\bibitem [{\citenamefont {Agrawal}(2012)}]{agrawal2012fiber}%
	\BibitemOpen
	\bibfield  {author} {\bibinfo {author} {\bibfnamefont {G.~P.}\ \bibnamefont
			{Agrawal}},\ }\href@noop {} {\emph {\bibinfo {title} {Fiber-optic
				communication systems}}}\ (\bibinfo  {publisher} {John Wiley \& Sons},\
	\bibinfo {year} {2012})\BibitemShut {NoStop}%
	\bibitem [{\citenamefont {Roy}\ \emph {et~al.}(2019)\citenamefont {Roy},
		\citenamefont {Mirajkar},\ and\ \citenamefont {Atkar}}]{14}%
	\BibitemOpen
	\bibfield  {author} {\bibinfo {author} {\bibfnamefont {A.}~\bibnamefont
			{Roy}}, \bibinfo {author} {\bibfnamefont {A.}~\bibnamefont {Mirajkar}},\ and\
		\bibinfo {author} {\bibfnamefont {G.}~\bibnamefont {Atkar}},\ }\bibfield
	{title} {\bibinfo {title} {Advancement on substitution cipher using multiple
			substitution table},\ }in\ \href@noop {} {\emph {\bibinfo {booktitle} {2019
				5th International Conference On Computing, Communication, Control And
				Automation (ICCUBEA)}}}\ (\bibinfo {organization} {IEEE},\ \bibinfo {year}
	{2019})\ pp.\ \bibinfo {pages} {1--6}\BibitemShut {NoStop}%
	\bibitem [{\citenamefont {Omran}\ \emph {et~al.}(2010)\citenamefont {Omran},
		\citenamefont {Al-Khalid},\ and\ \citenamefont {Al-Saady}}]{15}%
	\BibitemOpen
	\bibfield  {author} {\bibinfo {author} {\bibfnamefont {S.}~\bibnamefont
			{Omran}}, \bibinfo {author} {\bibfnamefont {A.}~\bibnamefont {Al-Khalid}},\
		and\ \bibinfo {author} {\bibfnamefont {D.}~\bibnamefont {Al-Saady}},\
	}\bibfield  {title} {\bibinfo {title} {Using genetic algorithm to break a
			mono-alphabetic substitution cipher},\ }in\ \href@noop {} {\emph {\bibinfo
			{booktitle} {2010 IEEE Conference on Open Systems (ICOS 2010)}}}\ (\bibinfo
	{organization} {IEEE},\ \bibinfo {year} {2010})\ pp.\ \bibinfo {pages}
	{63--67}\BibitemShut {NoStop}%
\end{thebibliography}
\end{document}